\newcommand{\COMMENT}[1]{}

\newcommand{\reff}[1]{}           % switch OFF SHOW REF in Bibliography, used in bibliograhy

\newcommand{\spaceline}[1]{\vspace{9pt}}
\newcommand{\spacex}[1]{\vspace{1cm}}
\newcommand{\spacexx}[1]{\vspace{2cm}}

\newcommand{\kl}[1]{\left\langle} % //or (
\newcommand{\kr}[1]{\right\rangle} % //or )

\newcommand{\BEGINFIGURE}{\begin{figure}[H] \center}         %%for the draft (ar arxiv), more pages
\newcommand{\NEW}[1]{#1}  %% OFF

      \newcommand{\MYTAB}[2]{
      \begingroup\parindent0mm\setlength{\leftskip}{#1} 
      #2 %% next empty line or \paris needed !!!
      \par  %\par (TeX) 	Ends the current paragraph. It is equivalent to leaving a blank line in the input. 
      \endgroup}
\documentclass[11pt,a4paper]{article}  %\documentclass[runningheads,a4paper]{llncs}                                       
\usepackage{multicol}  %used for CONTENTS

\usepackage[font=small,margin={9mm,9mm}]{caption} %used for smaller captions

\usepackage[margin=2.5cm]{geometry}

\usepackage{graphicx}

\usepackage{amssymb}
\usepackage{amsmath}

\usepackage[utf8]{inputenc}  %\usepackage[utf8]{inputenx}
\usepackage{url}             %% for special characters in email/url  -- necessary?

\usepackage[T1]{fontenc}

\usepackage{sidecap}   %sidecaption  --\begin{SCfigure} end..
\usepackage{here}      %for figure option [H]
\usepackage{setspace}  % http://ctan.org/pkg/setspace  RH change space between lines
\usepackage[colorlinks]{hyperref}
\begin{document}
%%%%%%%%%%%%%%%%%%%%%%%%%%%%%%%%%%%%%%%%%%%%%%%%%%%%
%%%%%%%%%%%%%%%%%%%%%%%%%%%%%%%%%%%%%%%%%%%%global settings RH
\setlength\parindent{0pt}
\setlength{\parskip}{6pt}       %between paragr.
\setlength{\overfullrule}{10pt}
%%%%%%%%%%%%%%%%%%%%%%%%%%%%%%%%%%%%%%%%%%%%global settings R

%%%%%%%%%%%%%%%%%%%%%%%%%%%%%%%%%%%%%%%%%%%%%%%%%%% paper TITLE and authors
\begin{center}
\Large
\textbf{Loop Patterns Formed by Cellular Automata}

	%~~\\
	%~~\\
	%\normalsize
	%\textit{Rolf Hoffmann}\footnote{
	%Technical University Darmstadt, Germany\\
	%hoffmann@informatik.tu-darmstadt.de}
	%%
	%and 
	%% %Mariusz Białecki
	%\textit{Mariusz Bia{\l}ecki}\footnote{
	%Institute of Geophysics Polish Academy of Sciences, Ksiecia Janusza 64, Warsaw, Poland\\~~~~~~ % ?? Księcia
	%bialecki@igf.edu.pl} 

~~\\
\large
\textit{Rolf Hoffmann}$^1$     and
\textit{Mariusz Bia{\l}ecki}$^2$               %Mariusz Białecki

\footnotesize
$^1$Technical University Darmstadt, Darmstadt, Germany\\
hoffmann@informatik.tu-darmstadt.de

%Ksi\k{e}cia   Księcia
$^2$Institute of Geophysics Polish Academy of Sciences, Ksi\k{e}cia Janusza 64, Warsaw, Poland\\  
bialecki@igf.edu.pl   
\normalsize           
\end{center}
%%%%%%%%%%%%%%%%%%%%%%%%%%%%%%%%%%%%%%%%%%%%%%%%%%% paper TITLE and authors
%%%%%%%%%%%%%%%%%%%%%%%%%%%%%%%%%%%%%%%%%%%%%%%%%%%%%%%%%%%%%%%%%%%%%%%%%%%%%%%% -----ARXIV header
%%%%%%%%%%%%%%%%%%%%%%%%%%%%%%%%%%%%%%%%%%%%%%%%%%%%%%%%%%%%%%%%%%%%%%%%%%%%%%%% -----ARXIV header

%%%%%%%%%%%%%%%%%%%%%%%%%%%%%%%%%%%%%%%%%%%%%%%%%%%%%%%%%%%%%%%%%%%%%%%%%%%%%%%% ARXIV abstract + kewords
%%%%%%%%%%%%%%%%%%%%%%%%%%%%%%%%%%%%%%%%%%%%%%%%%%%%%%%%%%%%%%%%%%%%%%%%%%%%%%%% ARXIV abstract + kewords
% __________________________________________________________________________ abstract
\begin{abstract}
\noindent
A Cellular Automata (CA) rule is presented that can generate
``loop patterns'' in a 2D grid under fixed boundary conditions.
A loop is a cyclically closed path represented by one-cells enclosed by zero-cells.
A loop pattern can contain several loops that are not allowed to touch each other.
The problem is solved by designing an appropriate set of tiles
that can overlap and which are used in the CA rule.
Templates are derived from the tiles which are used for local pattern matching.
In order to drive the evolution to the desired patterns,
noise is injected if the templates do not match or other constraints are not fulfilled.
The general CA rule can be specialized by enabling certain conditions,
and the characteristics of five rule variants are explained.
Simulations illustrate that the CA rule can securely evolve stable loop patterns.
The preliminary theoretical analysis of the obtained loop patterns raises many interesting research problems for the future
 --- several of them have been briefly discussed.
\end{abstract}
%Important Issues for Abstract: Objective:Method:Results:Conclusion: 
% __________________________________________________________________________ abstract

% __________________________________________________________________________keywords arxiv

% __________________________________________________________________________remark preprint
\noindent\textbf{Remark.} 
This article is a preprint version of:
\textit{Hoffmann, R. and Bia{\l}ecki, M.: 
Loop Patterns Formed by Cellular Automata}.
In: Adamatzky, A., Sirakoulis, G.C., Martinez, G.J. (eds) Advances in Cellular Automata. Emergence, Complexity and Computation, vol 52. Springer Cham (2025). 237--265.
\normalsize
% __________________________________________________________________________remark preprint
%%%%%%%%%%%%%%%%%%%%%%%%%%%%%%%%%%%%%%%%%%%%%%%%%%%%%%%%%%%%%%%%%%%%%%%%%%%%%%%% ARXIV abstract + kewords
%%%%%%%%%%%%%%%%%%%%%%%%%%%%%%%%%%%%%%%%%%%%%%%%%%%%%%%%%%%%%%%%%%%%%%%%%%%%%%%% ARXIV abstract + kewords

%__________________________________________________________________________ OPTION: table of contents
\vspace{3mm}
%~~\\
\hrule
\begin{multicols}{2}
%\newpage
\tableofcontents
\end{multicols}
\newpage
%__________________________________________________________________________ OPTION: table of contents

\normalsize
%%%%%%%%%%%%%%%%%%%%%%%%%%%%%%%%%%%%%%%%%%%%%%%%%%%%%%%%%%%%%%%% 
\section{Introduction}
\label{sec:Introduction}
%%%%%%%%%%%%%%%%%%%%%%%%%%%%%%%%%%%%%%%%%%%%%%%%%%%%%%%%%%%%%%%%

Loops are common in nature. They can be found in various organisms (e.g. veins in tree leaves, closed blood circulation system in animals), physical and geophysical systems (e.g. cracks on the surface of glazed ceramics or drying mud, tidal river deltas, highly porous limestone) or also in the products of human activity (e.g. street networks in cities). Such reticulated patterns (as opposed to tree patterns, where loops are absent) usually arise as a result of complex interactions, and understanding the mechanisms leading to their formation are a fascinating and challenging subject of research (see, e.g. \cite{A work on reticulated patterns}). On the other hand, patterns containing loops can be the subject of research in the field of constructing algorithms leading to their creation.

In this article, we investigate the formation of such loop patterns in which individual loops are separated from each other -- they do not intersect or even adhere to each other.
The aim in general is to generate such pattern classes by \textit{local conditions only} 
that are interesting from a global point of view.
An optional additional aim could be to optimize such patterns
with respect to a global measure, and if possible also in a local way.

%{\bf For instance, one could aim to fill a space with  particles where the
%distance between them should be kept in a certain range and their number should be 
%minimized or maximized.}
%%%%%%%%%%%%%%%%%%%%%%%%%%%%%%%%%%%%%%%%%%%%%%%%%%%%%%%%%%%%%%%%%%\subsection{Objectives}

%%%%%%%%%%%%%%%%%%%%%%%%%%%%%%%%%%%%%%%%%%%%%%%%%%%%%%%%%%%%%%%%%%
\subsection{The problem}
%%%%%%%%%%%%%%%%%%%%%%%%%%%%%%%%%%%%%%%%%%%%%%%%%%%%%%%%%%%%%%%%%%

The problem is to find \textit{loop patterns} in a 2D space 
by a Cellular Automata (CA) rule. 
\NEW{
We consider a square field/grid of cells of size $n \times n$  under
fixed to zero boundary conditions.
} 
First we  define what 
we mean by \textit{loop} and \textit{loop pattern}.

%%%%%%%%%%%%%%%%%%%%%%%%%%%%%%%%%%%%%%%%%%%%
\subsection{Loop and loop pattern}
\label{Loop and loop pattern}
%%%%%%%%%%%%%%%%%%%%%%%%%%%%%%%%%%%%%%%%%%%

\textbf{Loop.}
We define a \textit{loop} as a cyclically closed path on a background
(sequence of \textit{path cells}) without any intersections or connections. 
The value 1 is assigned to path cells (depicted in black/blue), 
and the value 0 to background cells (depicted in white/green). 
The loop length (the number of path cells) has to be greater than 1
(we forbid a single black cell to be a trivial loop). The shortest loop has the length 9.

A \textit{loop path cell} is a black cell that has exactly two black cells in NESW
where NESW are the direct orthogonal neighbors.
This means that three path cells in sequence form a vertical or horizontal straight line segment 
or a corner (4 variants by rotation). 
In order to form a loop it is required that 
the neighboring cells of a path cell are also path cells, and
all path cells in continuation form a closed path.

A loop path shall be surrounded by zero-cells (\textit{hull} cells).
This is already partly fulfilled by the definition
that each path cell has two white (just as two black) neighbors in NEWS.
Additionally, at each corner (convex side) an extra zero has to be provided
on the corner's outer diagonal cell, 
for example ~                              
$\substack{~~0~0
         \\0~1~1
         \\0~1~0}$  $\Rightarrow$ 
$\substack{0~0~0
         \\0~1~1
         \\0~1~0}$~~. %	~where '.' marks an extra zero.  
The hull ensures that loops keep a (visible) minimal distance
between each other, and also within a loop itself.
Hull cells are allowed to be used by several paths,
from the same or other loops.

\NEW{
In summary, 
a path of one-cells is closed and it is embedded in a zero-hull
if
each one-cell fulfills the following two conditions:
(i) it has two one-cells in NESW, 
and 
(ii)  every corner's outer-diagonal one-cell is zero.
We call these two conditions in conjunction the  ``loop path condition''.
}

\textit{Remark.} We may imagine that an agent walks around on the grid in one-way
and returns back to its starting point without touching     %(keeping distance of at least one white cell)
its own path (except for the cyclic connection) or another loop.
The agent follows its path by going straight, to the right or to the left.
When walking along the path, the visited path cells are colored to black,
assuming that the field was initially colored in white. 
%assuming that the agent has a direction. 

\textbf{Loop pattern.}
A \textit{loop pattern} is a cell field with a white background 
that contains at least one loop.
Several loops of different types may appear in such a pattern,
and a loop may enclose other loops.
Note that connected and intersecting loops are not allowed.
The field size $n$ can be arbitrary, but we have to notice that 
not every loop type can appear for an arbitrary size. 
For instance the smallest  $3\times3$ loop (Fig.~\ref{u4x4-1})
needs at least a space of $5\times 5$ including the zero border cells. 
Therefore $n\geq3$ because  the minimal size of the active cell area is $3\times 3$.

\textit{Remark.} An \textit{optimal loop pattern} is a loop pattern that 
maximizes or minimizes a global criterion.
A global criterion is a measure 
that may take all cell states into account.
Usually it is easily computed on the global level, often by counting some 
local conditions or other relations between arbitrary cells. 
For example one could aim at \textit{dense} loop patterns
in which the number of path cells is maximal. 
In this work we will not aim at optimal patterns.  

%%%%%%%%%%%%%%%%%%%%%%%%%%%%%%%%%%%%%%%%%%%%%%%%%%
\subsection{How the problem was solved}
%%%%%%%%%%%%%%%%%%%%%%%%%%%%%%%%%%%%%%%%%%%%%%%%%%

The main challenge is how to form a loop (its path and closing it) by
local conditions.
The solution presented here is the definition of a set of tiles 
(local patterns acting as local conditions) that 
are allowed and forced to overlap (aggregate) in a way
that loops are formed.
Such a set of tiles was designed and tested by hand and
by a special tool 
\cite{2023-Hoffmann-mdpi-loop-pattern}
based on a Genetic Algorithm. It can produce 
patterns for a given set of tiles taking into account a global optimization criterion. 

After having found an appropriate set of tiles, the second challenge
is to find a \textit{local} Cellular Automata (CA) rule that can evolve the desired loop patterns.
To solve this problem, we extend the approach already used to generate
point patterns 
\cite{2022-2019-Arxiv-Forming Point Patterns by a Probabilistic Cellular Automata Rule}, %1
domino patterns 
\cite{Hoffmann:Deserable-Seredynski-2021a-JSup-A cellular automata rule placing a maximal number of dominoes in the square and diamond}, 
sensor networks
\cite{Hoffmann:Deserable-Seredynski-2022-NatCom-Cellular automata rules solving the wireless sensor network coverage problem},
and loop patterns on the torus
\cite{2023-Hoffmann-mdpi-loop-pattern}.
Local matching patterns (templates) are derived from the tiles and
tested by the CA rule. If there is no template hit, noise is injected 
that finally drives the evolution to a loop pattern. 
%%%%%%%%%%%%%%%%%%%%%%%%%%%%%%%%%%%%%%%%%%%%%%%%%%%%%%%%%%%%%%%%%%\subsection{Applications}

%%%%%%%%%%%%%%%%%%%%%%%%%%%%%%%%%%%%%%%%%%%%%%%%%%%%%%%%%%%%%%%%%%
\subsection{Related work}
%%%%%%%%%%%%%%%%%%%%%%%%%%%%%%%%%%%%%%%%%%%%%%%%%%%%%%%%%%%%%%%%%%
% related work (currents state, citing key publications), continuation of the work

\textit{Own Previous Work.}
Complex pattern were constructed in a local way by moving CA agents
\cite{2014-Hoffmann-ACRI-How Agents Can Form a Specific Pattern}. %5
%\cite{2016-Hoffmann-Polonia-Cellular Automata Agents form Path Patterns Effectively}%6
%\cite{2016-Hoffmann-D-ACRI-Line Patterns Formed by Cellular Automata Agents}%7
%\cite{2017-Hoffmann-D-PACT-Generating Maximal Domino Patterns by Cellular Automata Agents}. %8
%\cite{Hoffmann:Deserable-2019-Domino pattern formation by cellular automata agents}. %9
%
The agents use a finite state machine evolved by a Genetic Algorithm (GA). 
Though the results are impressive, it takes quite some effort to train the agents. 
In addition it is very difficult to train them for small and large grid sizes at the same time
in order to make them work for any grid size.
It would be a new interesting task to train them for forming loop patterns
and to find out the limitations.  

Then a probabilistic CA rule was used to generate point patterns
\cite{2022-2019-Arxiv-Forming Point Patterns by a Probabilistic Cellular Automata Rule}, %1,
sensor point patterns
%
%\cite{Hoffmann:Seredynski-2021-ACRI-Covering the space with sensor tiles}%10
\cite{Hoffmann:Deserable-Seredynski-2022-NatCom-Cellular automata rules solving the wireless sensor network coverage problem}, 
and domino patterns
%\cite{2019-pact-A Probabilistic Cellular Automata Rule Forming Domino Patterns}%2
\cite{Hoffmann:Deserable-Seredynski-2021a-JSup-A cellular automata rule placing a maximal number of dominoes in the square and diamond}%3
\cite{Hoffmann:Deserable-Seredynski-pact-2021b-Minimal Covering of the Space by Domino Tiles}. % 11
It was possible to aim at a maximal or a minimal number of dominoes by injecting noise depending on the level of tile overlapping. 
In this paper the idea of using a probabilistic CA is also followed. 
The task of generating loop pattern under cyclic boundary condition was already addressed in our last work
\cite{2023-Hoffmann-mdpi-loop-pattern}.
There more complicated tiles were used that may lead to faulty patterns which could only be avoided by an additional condition.
Now we found a more suitable and simpler set of tiles that directly produce non-faulty loop patterns 
even for the more complicated case with fixed boundaries. 
%Now the problem is more complicated because of the fixed boundaries, and we could found a more simple set of tiles.
%The rule here was adapted to a new set of tiles and uses an additional logical condition in order to evolve loop patterns \textit{only}.

\textit{Overlapping Tiles and Loops.} 
The recommendable book \cite{1987 Tilings} provides much information about tilings but it does not 
deal with overlapping tiles. 
Overlapping 1D tilings are treated theoretically in the context of recognizable languages, monoids, and
two-way automata
%
%\cite{2013 Janin On languages of one-dimensional overlapping tiles}.
%Janin, D. (2013). On languages of one-dimensional overlapping tiles. In SOFSEM 2013: Theory and Practice of Computer Science: 39th International Conference on Current Trends in Theory and Practice of Computer Science, Špindlerův Mlýn, Czech Republic, January 26-31, 2013. Proceedings 39 (pp. 244-256). Springer Berlin Heidelberg.
%
\cite{2013 Janin On languages of one-dimensional overlapping tiles}, 
and it was partly motivated by an application in computational music theory. 
This paper is a good basis for further studies and extensions to the $n$-dimensional case.

There are a lot of applications for overlapping tilings, like the dense parking of cars in a parking lot,
or constructing a sieve for maximal particle flow
(using domino tiles) \cite{Hoffmann:Deserable-Seredynski-pact-2021b-Minimal Covering of the Space by Domino Tiles},
or optimizing task scheduling (overlapping communication and computation)
%
%\cite{2001 Minimizing completion time for loop tiling with computation and communication overlapping}Goumas, G., Sotiropoulos, A., & Koziris, N. (2001, April). Minimizing completion time for loop tiling with computation and communication overlapping. In Proceedings 15th International Parallel and Distributed Processing Symposium. IPDPS 2001 (pp. 10-pp). IEEE.
\cite{2001 Minimizing completion time for loop tiling with computation and communication overlapping},
or building nano-structures
\cite{2001 Nanoparticles}. 
%Waychunas, G. A. . Structure, aggregation and characterization of nanoparticles. 
%Reviews in Mineralogy and Geochemistry, 44(1), 105-166.
%(2001)

Note that in our approach the space between the paths of the loop is 
not constant 1 (assuming a grid solution), it can vary between 1 and 2 (or even more if uncovered cells are allowed).
Thus our loops can be arranged in a more flexible way for 
any grid size, loop paths can keep a suitable distance out of a certain range. 
%{\bf This feature is also valuable for describing particles that can attract and repulse.}

Our problem is related to closed (loop) \textit{space-filling curves}
\cite{1890Peano}
\cite{1891Hilbert}
\cite{1991SpaceFilling}.
%
%%%%%%%%%%%%%%%%%%%%%%%%%%%%%%%%%%%%%%%%%%%
%\bibitem{1890Peano}
%Peano, G.
%Sur une courbe, qui remplit toute une aire plane.
%\textit{Mathematische Annalen (in French), 36 (1): 157–160,}
%\textbf{(1890)}
%\bibitem{1991SpaceFilling}.
%Prusinkiewicz, P.; Lindenmayer, A.; Fracccia, D.
%Synthesis of space-filling curves on the square grid.
%1989
%In 
%Fractals in the Fundamental and Applied Sciences,
%Peitgen, H.-O.; Henrique, J.M.; Pencdo, L.F. (Editors),
%Elsevier Science Publishers B.V. (North-Holland)
%1991.
%
%\bibitem{1891Hilbert}
%Hilbert, D.
%Ueber die stetige Abbildung einer Linie auf ein Flächenstück.
%\textit{Mathematische Annalen (in German), 38 (3): 459–460}
%\textbf{(1891)}
%Przemyslaw Prusinkiewicz, Aristid Lindenmayer, and F. David Fracchia. 1989.
%SYNTHESIS OF SPACE-FILLING CURVES ON THE SQUARE GRID
%Przemyslaw Prusinkiewiczl, Aristid Lindenmayert, and F. David Fracchia
%in
%Fractals in the Fundamental and Applied Sciences
%H.-O. Peitgen, J.M. Henriques & L.F. Pencdo (Editors)
%Elsevier Science Publishers B.V. (Nortl1-Holland)
%1991
%Peano, G. (1890), "Sur une courbe, qui remplit toute une aire plane", Mathematische Annalen (in French), 36 (1): 157–160
%%%%%%%%%%%%%%%%%%%%%%%%%%%%%%%%%%%%%%%%%%%%%%%%%%%%%%%%%%%%%%%
%
In principle such loops can be generated with the presented CA approach, among other techniques.
(An example is the loop shown in Fig.~\ref{ALL7x7} a1.)
Generating \textit{only} space-filling curves by a CA rule is a future work. 
For space-filling curves
% the dependency of possible solutions on the field size $n$ has to be considered.
the field size  has to be considered with respect to the possible solutions.
Hamiltonian cycles (i.e. cycle that visits each site exactly once) can algorithmically be computed in grid graphs.
In \cite{CrossleyHamiltonian} the grid graph is interpreted as a field of cells as here,
whereas in  
\cite{2012 Hamiltonian}
a checkerboard marked grid graph is used. 
\COMMENT{
\bibitem{CrossleyHamiltonian}
Crossley, M.
Hamiltonian Cycles in Two Dimensional Lattices,
\textit{Statistical Physics in Biology,
Center for Theoretical Physics, MIT}
\textbf{(19xx).}

\bibitem{2012 Hamiltonian}
Keshavarz-Kohjerdi, F.; Bagheri, A. 
Hamiltonian paths in some classes of grid graphs. 
\textit{Journal of Applied Mathematics.}
\textbf{(2012).}
}
In
\cite{1999StringTopology} 
families of intersecting closed curves (connected loops) are discussed in the context
of string topologies. This paper could be of interest for a further 
work on generating connected loops.

%%%In graph theory, a ``loop'' (also called a self-loop) is an edge that connects a vertex to itself,and a ´´cycle'' corresponds to a loop in our context.

\COMMENT{
\bibitem{1999StringTopology} 
Chas, M.;  Sullivan, D. 
String topology.
arXiv preprint math/9911159.
(1999).
} 

%-------------------------------------------------------
\textit{Cellular Automata.}
A CA is a grid of automata (cells).
Every cell changes its state depending on its own state and the state of its local neighbors
according to a local rule. 
The model of computation is
parallel, simple, powerful and has wide range of applications 
%\cite{2002 Wolfram}
%Wolfram, S. (2002). A new kind of science (Vol. 5, p. 130). Champaign: Wolfram media.
\cite{2002 Wolfram}.

\textit{Paper organization.}
%This chapter is organized as follows.
In the next Sect. 
\ref{S2}
we present the concept of \textit{overlapping tiles} that can form a pattern.
In the main Sect.  \ref{S3}
the CA rule and its simulation is presented. 
A set of overlapping tiles is defined that can aggregate to loops. 
From each tile several \textit{templates} (shifted tiles) are derived to be used in the CA rule. 
Then the general CA rule is given, and five variants %specializations 
thereof (\textit{Rule0} to \textit{Rule4}) are described.
These specialized rules generate loop patterns with different properties. 
In Sect. \ref{S4}
the CA rules are simulated and the evolved patterns are discussed for different field sizes. 
Then in Sect. \ref{S5}
this article is summarized and
open issues for future work are presented. 
%%%%%%%%%%%%%%%%%%%%%%%%%%%%%%%%%%%%%%%%%%%%%%%%%%%%%%%%%%%%%%%%%%\subsection{Related Work}

%%%%%%%%%%%%%%%%%%%%%%%%%%%%%%%%%%%%%%%%%%%%%%%%%%%%%%%%%%%%%%%%---- Introdution
%==============================================================new intro

%%%%%%%%%%%%%%%%%%%%%%%%%%%%%%%%%%%%%%%%%%%%%%%%%%%%%%%%%%%%%%%%
\section{Overlapping tiles}
\label{S2}
%%%%%%%%%%%%%%%%%%%%%%%%%%%%%%%%%%%%%%%%%%%%%%%%%%%%%%%%%%%%%%%%
In this section the notion of \textit{tiles} and their \textit{overlapping} is introduced.

\textbf{Tile.}
%A \textit{tile support} $S$ is a pattern  of $m_1 \times m_2$ elements $L(x,y) \in\{0,1,\textit{-}\}$ where $m_1, m_2 \leq n$. 
A \textit{tile support} $S$ is a pattern  of $m_1 \times m_2$ elements $L(x,y) \in\{0,1,-\}$ where $m_1, m_2 \leq n$. 
We call the elements ``pixels'' in order to distinguish them from cells. 
A \textit{tile} $L$ is a pattern of all elements $\{0,1\}$ of the tile support $S$. 
The symbol ``-'' represents a \textit{null}, a pixel that is not part of a tile.
A non-null pixel 0/1 is also called a \textit{valid pixel}. 
We use only tiles that are much smaller (or smaller for small fields) than the size of the whole field.
The reason is that we do want to allow that a few large tiles of size $m_1 \approx n$ and/or $m_2 \approx n$ may simply cover the whole global space.
Then, a trivial solution could be that just one tile is equal to the aimed pattern.
In fact, we want to restrict the size of the tiles,  $m_1 \leq B_{m_1}$ and $m_2 \leq B_{m_2}$, where
$B_{m_1}, B_{m_2} \geq 1$  are the boundaries of a tile,
and where both $B_{m_1}$ and $ B_{m_2} \geq 2$.
Here, we want to solve the problem with small tiles of size $B_{m_1},B_{m_2} \leq 3$. 
The coordinates $(x,y)$ within a tile $L$ are locally defined and separately for each tile where its \textit{origin} (or its \textit{anchor}) $(0,0)$ 
is assigned to the central pixel of the tile, in a case of both $ m_1$ and $m_2$ are odd numbers, otherwise to a (chosen)  pixel adjacent to the central point of the tile support.

\textit{{Remark.}}
It would be possible to define a tile as a set of pixels $(x,y,v)$ where 
$(x,y)$ are the local coordinates, and $v\in\{0,1\}$ are the pixels' states. In addition, the name of the tile and the anchor's position have to be related. 
%This is a good option to describe In the literature \ref{}+++Grünbaum, B. and Shephard, G. C. Tilings and Patterns. New York: W. H. Freeman, 1986.a tiles is usually defined as a set of points. 

\textbf{Overlapping.}
Now, we search for global patterns that can be tiled by tiles from a set of tiles
$\underline{L} = \{L_0, L_1,$ \dots.\} that may overlap.
Two tiles $L1, L2 \in \underline{L}$ , $L1 \neq L2$, \textit{{overlap}} 
if they are overlaid (shifted relative to each other),
and all overlaid valid pixels have the same value. 
The result of overlaying a pixel $p_1$ from $L_1$ 
with a pixel $p_2$ from $L_2$
is summarized in Table~\ref{TablePixel}. 
	
%COMMENT alternatively it could be written in some formal notation like 
%overlay(p1,p2)=overlay(p2,p1)
%overlay(p1,p2)=a if p1=p2=a in {0,1}
%overlay(p1,p2)=b if p1='-' and p2 in {0,1}
%overlay(0,1) is a forbidden conflict
%overlay(-,-)='-' is not considered as an overlay

%_______________________________________________________________
	\begin{table}[h]
  \center
	\caption{The result of overlaying two pixels.}
	
	\begin{tabular}{cc|c}
		           pixel $p_1$    &  pixel $p_2$	  &   result     \\
  \hline                                                 
	             0           &  0           &  0           \\
							 1           &  1           &  1           \\
							 -           &  0/1         &  0/1           \\
							 0/1         &  -           &  0/1           \\
               0           &  1           &  forbidden   \\
	             1           &  0           &  forbidden   \\
 
  \end{tabular}
	\label{TablePixel}
	\end{table}
	%_______________________________________________________________

It is possible that $k\geq 2$ tiles overlap. In this case,  
there exists pairwise an overlap and the resulting overlap forms a 
\textit{{compound tile}} (or \textit{macro-tile})
consisting of $k$ sub-tiles.
A \textit{compound tile} is an assembling of basic tiles or other compound tiles. 
In order to cover a given field, not only the basic tiles but also compound tiles can be used. 
%
%In terms of particle systems, we may associate a basic particle with a (basic) tile, and a compound/complex particle with a compound tile.%Here we want to construct patterns by assembling and overlaying tiles. 

We define the \textit{cover level} $v$ as the number of overlaid valid pixels at a certain site.
For an uncovered site / cell  yields $v=0$, and there is a maximum $v_{max}$ depending on the 
given set of tiles and the actual aggregation / pattern. 
See for example the overlapping tiling patterns in Fig.~\ref{GraphTilingLast}
where $v_{max}=8$ for the $3\times3$ loop.
We define a field of cells as \textit{{fully covered}} if all cells of a given field are covered 
by at least one tile pixel (i.e. a valid pixel)
taken from the given set of tiles that are allowed to overlap.
Due to a not properly defined set of tiles,
it may happen that a given field cannot totally be covered by valid pixels.
Then, the field is \textit{{partially}} (not fully) covered because it contains uncovered cells (gaps, holes, spaces). 
Nevertheless, partially covered fields can be of interest as patterns, too.

%%%%%%%%%%%%%%%%%%%%%%%%%%%%%%%%%%%%%%%%%%%%%%%%%%%%%%
\section{The simulation and design of the CA rule}
\label{S3}
%%%%%%%%%%%%%%%%%%%%%%%%%%%%%%%%%%%%%%%%%%%%%%%%%%%%%%

In the following 
Sections
~\ref{Updating} and
Sect.~\ref{The Simulation Algorithm}
we explain the used asynchronous updating scheme and how the CA rule is simulated. 
Then the tiles are defined 
(Sect.~\ref{The Defined Tiles}) 
from which the templates are derived
(Sect.~\ref{Templates Derived From the Tiles}),
and then in 
Sect.~\ref{The Rule Details}, 
the rule with its variants (\textit{Rule0} to \textit{Rule4}) is presented  in detail.
The rules inject noise as long as no loop pattern is reached.
Simulations follow showing the resulting patterns. 

%%%%%%%%%%%%%%%%%%%%%%%%%%%%%%%%%%%%%%%%%%%%%%%%%%%%%%
\subsection{Chosen updating scheme}
\label{Updating}
%%%%%%%%%%%%%%%%%%%%%%%%%%%%%%%%%%%%%%%%%%%%%%%%%%%%%%

The simulation of a CA is very simple, that is one reason why CA are so popular. 
Nevertheless, there are different types of CA 
and we need to choose the best-suited one for our application.
There are two important properties we have to choose:
(i) \textit{{synchronous}} or \textit{{asynchronous}} updating, 
and (ii) a \textit{{deterministic}} or a \textit{{probabilistic}} rule.

\vspace{9pt}
\textbf{(i) Synchronous or asynchronous updating.}

\begin{itemize}
\item 
\textit{{Synchronous updating.}} 
	
(Phase 1) For every cell $(x,y)$ its new state value $s'$ is computed by a local rule
and buffered in the \textit{new state} variable $s_{new}$. 

\vspace{5pt}
%% $\forall{(x,y)}: s_{new}(x,y) \leftarrow s'(x,y) = f(s(x,y), s(neighbor_1(x,y)), s(neighbor_2(x,y)), \ldots)$

~~~$\forall{(x,y)}: s_{new}(x,y) \leftarrow s'(x,y) = f(\textit{NeighborhoodStates}(x,y))$

where

~~~$\textit{NeighborhoodStates}(x,y)=(s(x,y), s(nbr_1(x,y)), s(nbr_2(x,y)), \ldots)$

~~~~~~where $nbr_i$ denotes a local neighbor. 

\vspace{5pt}
We use the ``arrow'' assignment  $z\leftarrow v$ to denote that  $z$ is a variable with a memory  which  stores the value $v$.

\vspace{5pt}
(Phase 2) The state of each cell is updated (replaced) by its new~state.

\vspace{5pt}
~~~$\forall{(x,y)}: s(x,y) \leftarrow s_{new}(x,y)$  %for every $(x,y)$. 

\vspace{6pt}
It is important to notice that the order in which the cells
are processed in phase 1 and in phase 2 does not matter,
but the phases must be separated.
This model can easily be implemented in software,
 or in clocked hardware using d-flipflops (internally supplying a master memory $s_{new}$ and a slave memory $s$).

\vspace{5pt}

\item
\textit{{Asynchronous updating.}}

There are several schemes.
In the pure case, there are no phases.
A cell $(x,y)$ is selected at random, 
its new state value $s'$
is computed,
and then immediately used for updating the cell's state: 
$s \leftarrow s'$.

\vspace{6pt}
We use the following scheme.
A time step 
$t \rightarrow t+1$  %$\Delta t$ 
is considered as a block of $N=n \times n$
micro time steps $\tau \rightarrow \tau +1$.
The selected cell is processed during one micro time step. 
For each time step, we select the cells sequentially in a 
random order. 
During one time step, each cell is updated once, but~the order is random. 
We display  CA configurations at time steps. 
This scheme is called \textit{{random sequence}} or \textit{{random new sweep}}. 
\end{itemize}

%\noindent\textbf{{Deterministic or Probabilistic Rule.}}
\textbf{(ii) Deterministic or probabilistic rule.}
A \textit{{deterministic}} rule always computes the same next state for 
a given input (the combined state of all neighbors). 
A \textit{{probabilistic}} rule computes
different next states with certain associated probabilities for a given input.

%\noindent\textbf{{Combination.}}
\textbf{Combination of (i) and (ii).}
In principle, we can combine
\textit{{synchronous}} or  \textit{{asynchronous}} updating
with 
a \textit{{deterministic}} or a \textit{{probabilistic}} rule.
This makes four options:
(1) synchronous updating and deterministic rule;
(2) synchronous updating and probabilistic rule;
(3) asynchronous updating and deterministic rule;
(4) asynchronous updating and probabilistic rule.

\textit{{Option 1:}} 
Until now, it was not possible to design %such a rule.
such a rule that can produce loop patterns.
The problem is that the evolving patterns may get stuck in non-desired patterns or
oscillating structures such as we know from the \textit{{Game of Life}}.
It remains an open question if it is possible to find such a rule.

\textit{\textit{{Options 2 -- 4:}}}
These options are related because the computation of a new configuration is stochastic. It seems that they can be transformed into each other to a certain extent. 
We decided  to use option 4 (random sequence together with a probabilistic rule)
as in  our former work 
%
%\cite{2022-2019-Arxiv-Forming Point Patterns by a Probabilistic Cellular Automata Rule,Hoffmann:Deserable-Seredynski-2021a-JSup-A cellular automata rule placing a maximal number of dominoes in the square and diamond,Hoffmann:Deserable-Seredynski-pact-2021b-Minimal Covering of the Space by Domino Tiles,Hoffmann:Deserable-Seredynski-2022-NatCom-Cellular automata rules solving the wireless sensor network coverage problem,2023-Hoffmann-mdpi-loop-pattern}.
%
\cite{2023-Hoffmann-mdpi-loop-pattern}
\cite{2022-2019-Arxiv-Forming Point Patterns by a Probabilistic Cellular Automata Rule}
\cite{Hoffmann:Deserable-Seredynski-2021a-JSup-A cellular automata rule placing a maximal number of dominoes in the square and diamond}
\cite{Hoffmann:Deserable-Seredynski-2022-NatCom-Cellular automata rules solving the wireless sensor network coverage problem}
\cite{Hoffmann:Deserable-Seredynski-pact-2021b-Minimal Covering of the Space by Domino Tiles}.

%
%\cite{2019-pact-A Probabilistic Cellular Automata Rule Forming Domino Patterns}
%\bibitem{2019-pact-A Probabilistic Cellular Automata Rule Forming Domino Patterns}
%\bibitem{2022-2019-Arxiv-Forming Point Patterns by a Probabilistic Cellular Automata Rule}
%
With asynchronous updating, we do not need buffered storage elements and a central clock for synchronization, which is closer to the modeling of natural~processes.

\newpage
%%%%%%%%%%%%%%%%%%%%%%%%%%%%%%%%%%%%%%%%%%%%%%%%%%%%%%%%%
\subsection{The simulation algorithm}
\label{The Simulation Algorithm}
%%%%%%%%%%%%%%%%%%%%%%%%%%%%%%%%%%%%%%%%%%%%%%%%%%%%%%%%

The simulation algorithm uses the following data elements.

\begin{itemize}
	\item 
	$q=(s,s',h_0,h_1,v)$  
	
	~~is the \textit{full cell state}, where %? full -> compound

	\begin{itemize}
	\item [] 
		$s\in\{0,1\}$  
		
		~~is the \textit{state} or \textit{pattern state}.
				%$The border value is -1 and the color is 0,1.

	\item [] 
	$s'\in\{0,1\}$ 
									
	~~is a temporary variable holding the new state value.

	\item [] 	
	$h_0\in\{0 \ldots 8\}$           
	
	~~is the number of hits if s'=0.

	\item [] 
	$h_1\in\{0 \ldots 3\}$           
	
	~~is the number of hits if s'=1. 				
		
	\item [] 
	$v\in\{0 \ldots 8\}$           
	  
  \MYTAB{3mm}  % equals to ~~ 
  {is the cover level. It is used to show how many valid pixels are overlaid.
	For the functionality to evolve a loop pattern it is not needed. (Note, that the maximal values of $h_0, h_1, v$ are given for the set of tiles shown in Fig. \ref{TilesLoopUnconnected}. For another set they can be different.)}
  
	\end{itemize}	
	
	\item 	
	$c=\textit{array}[0 \ldots n-1, 0 \ldots n-1] ~\textit{of} ~q$    	
  
  \MYTAB{3mm}  % equals to ~~
  {is the field of active cells (except borders). $c[x,y]$ denotes a cell at the site $(x,y)$. 
	The dot-notation can be used to specify an element of the compound state, e.g.
	$c[x,y].s = s[x,y]$ is the pattern state at $(x,y)$.}	
	
  \item
  $Z=((x,y)), ~~x=0 \ldots n-1, y=0 \ldots n-1$
  
  ~~is an index vector defining a sequence of coordinates for selecting a cell.
\end{itemize}

The algorithm (Fig.~\ref{simalgorithm}) describes how the CA system is simulated
under asynchronous updating. 
The used CA rule is described in detail in the next sections.  
At each time-step all cells are selected according to the random order given by $Z$,
and then the CA rule is applied to the selected cells.

%%%%%%%%%%%%%%%%%%%%%%%%%%%%%%%%%%%%%%%%% simalgorithm
%\begin{figure}[H]
\BEGINFIGURE
\hrulefill
%\rule{\textwidth}{0.5pt}
\begin{itemize}%%%%%%%%%%%%%%%%%%%%%%%%%%%%%%%%%%%%%%%%%%%%%%%%%%%%%%%%%%%%% 1
	\item[I] 
	initialization

  \begin{itemize}
    \item [1:]  $s[x,y] \leftarrow \textit{random} \in\{0,1\}, h_0=h_1=0, ~\forall (x,y)$ in active area 
    \item [2:] $t\leftarrow 0$
    \item [3:] $Z \leftarrow XY$     \hfill-- array of cell indices
  \end{itemize}

	\item[II]
	\textbf{repeat}
	    
			\begin{itemize}%%%%%%%%%%%%%%%%%%%%%%%%%%%%%%%%%%%%%%%%%%%% 2
				\item[1:] 
			  compute tile matches and cover levels, and then show pattern
        
        \item[2:] 
        \textbf{exit} repeat loop \textbf{if} $t=T_{max}$  \hfill-- termination of the simulation      
				
				\item[3:]
				compute next generation

        \begin{itemize}%%%%%%%%%%%%%%%%%%%%%%%%%%%%%%%%% 3
           \item[a:]    $t\leftarrow t+1$  \hfill-- increment time-counter / generation
                      
           \item[b:]   $Z \leftarrow \textit{Permute}(Z)$  \hfill-- make a new random sequence of indices         
          
           \item[c:]   \textbf{foreach} $(i,j)$ \textbf{in} $Z$ \textbf{do} \hfill-- each cell of the field is processed                  
                   
                   \begin{itemize}%%%%%%%%%%%%%%%%%%% 4
                     \item[1:]   $(x,y)\leftarrow Z[i,j]$   \hfill-- select a new cell                    
                    
                     \item[2:]   $s[x,y] \leftarrow \textit{CARule}(\textit{Neighborhood}(x,y))$  \hfill--   compute $s',h_0,h_1$ and update state                         
                   \end{itemize}%%%%%%%%%%%%%%%%%%%% 4
            
         \end{itemize}%%%%%%%%%%%%%%%%%%%%%%%%%%%%%%%%% 3  
            
			\end{itemize}%%%%%%%%%%%%%%%%%%%%%%%%%%%%%%%%%%%%%%%%%%%%%%%% 2

\end{itemize}%%%%%%%%%%%%%%%%%%%%%%%%%%%%%%%%%%%%%%%%%%%%%%%%%%%%%%%%%%%%% 1
\hrulefill
%\rule{\textwidth}{0.5pt}
\caption{%++simalgorithm++
This algorithm describes how  the CA rule is asynchronously applied.
(3:) For each time-step 
(c:) every cell is selected according  to the random order given by $Z$.
(c1:) A cell $(x,y)$ is selected, and
(c2:) the CA rule is applied.
}
\label{simalgorithm}   
\end{figure}
%%%%%%%%%%%%%%%%%%%%%%%%%%%%%%%%%%%%%%%%% simalgorithm

(I)
Initially the state $s$ is random, if not specified otherwise. 
The template hits $h_0, h_1$ (described in Sect.~\ref{The Rule Details} are set to zero, but 
they could be set to other values to influence the evolution.
They could also be set to the initial cover level
in order to immediately stabilize a given initial loop pattern.
The borders are constantly set to the value $s=0$.
The active area is the inner $n\times n$ area where the rule is applied.
A certain order of the $(x,y)$-coordinates is initially assigned to $Z$.

(II) 
This loop is repeated until the exit condition (2:) is fulfilled. 
The first step (1:) is optional but very useful.
Each tile of the tile set is checked at every $(x,y)$ if it matches
with respect to the current local pattern states $s$.
In case of a match the cover levels $v$ are updated.
The match positions
where certain tiles match (their centers) can separately be stored
for statistics and visualization purposes (the little dots in figures like
Fig.~\ref{TilesLoopUnconnected} or
Fig.~\ref{ALL5x5}).

%%%%%%%%%%%%%%%%%%%%%%%%%%%%%%%%%%%%%%%%%%%%%%%%%%%%%%
\subsection{The defined tiles}
\label{The Defined Tiles}
%%%%%%%%%%%%%%%%%%%%%%%%%%%%%%%%%%%%%%%%%%%%%%%%%%%%%%

The designed set of tiles which can produce loop patterns is shown in 
Fig.~\ref{TilesLoopUnconnected}.
Each tile has a center (marked dot) and contains three consecutive one-pixels in order to form a path
by overlapping. 
The corner tiles are used to form corners and the line tiles are used to form lines (horizontally or vertically).
In Fig.~\ref{GraphTilingLast} (I)
we can see how a $3\times3$ square loop can be build. 
Starting with an empty field (a) 
the upper left corner tile A0 is placed (b),
then the upper right corner A1 is placed (c),
then a horizontal line segment B0 is placed between them (d),
and then the tiles B1, A2, B0, A3, B1 are placed in continuation. 
To each cell of the loop a tile is linked with its center,
and the neighboring tiles overlap
 such that the cover level 
of each path cell yields $v=3$.
\NEW{
Thereby the \textit{loop path condition} is realized.
}
%(the path is closed and enclosed by zeroes, 
%(Sect. \ref{Loop and loop pattern})
%
It is obvious that the order in which the tiles are aggregated for a certain pattern does not matter.
Nevertheless does the order in which tiles are placed influence  
the further transient or final pattern.
%

%===================================================================================== TilesLoopUnconnected
\vspace{-3pt}
%\begin{figure}[H]             %10.5 cm  %\includegraphics[width=3cm]{Figures/grafik-1-h1-nr3}
\BEGINFIGURE
(I) 

~~~~~~~\includegraphics[width=10cm]{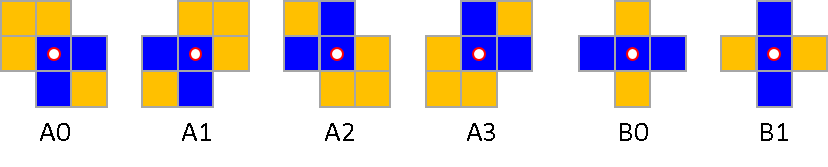} 

(II) 

~~~~~~~\includegraphics[width=10cm]{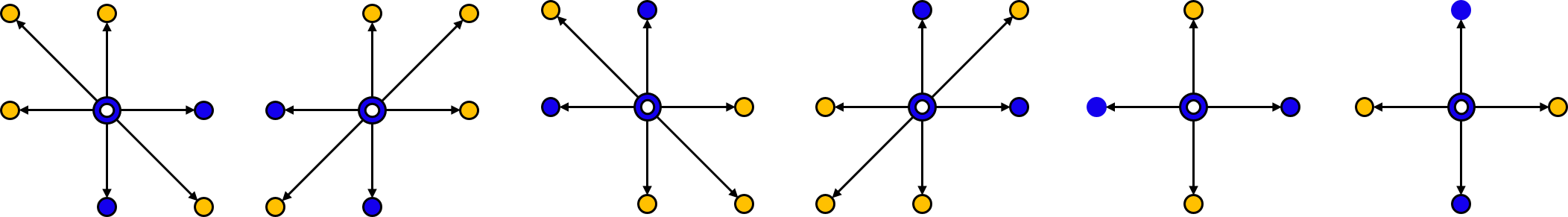}
\caption{%+++Figure TilesLoopUnconnected+++
The defined set of tiles.
(I) The corner tiles $A0, A1, A2, A3$ and the line tiles $B0, B1$. 
(II) In the second row the tiles are represented as graphs,
in which the valid tile pixels correspond to vertices, and the center vertex
has outgoing edges to the remaining vertices.
}
\label{TilesLoopUnconnected}
\end{figure}   
\unskip 
\vspace{11pt}
%===================================================================================== TilesLoopUnconnected

%===================================================================================== GraphTilingLast
\vspace{-3pt}
\BEGINFIGURE
(I)

(a) \includegraphics[width=0.15\textwidth]{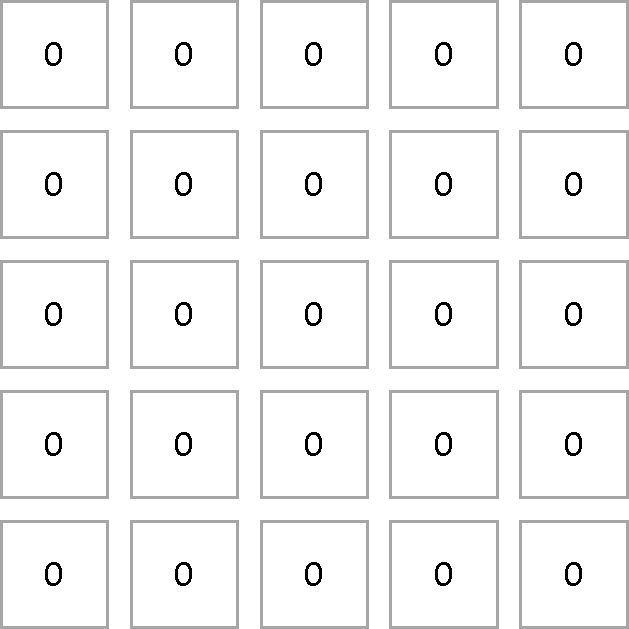}   %$\Rightarrow$
(b) \includegraphics[width=0.15\textwidth]{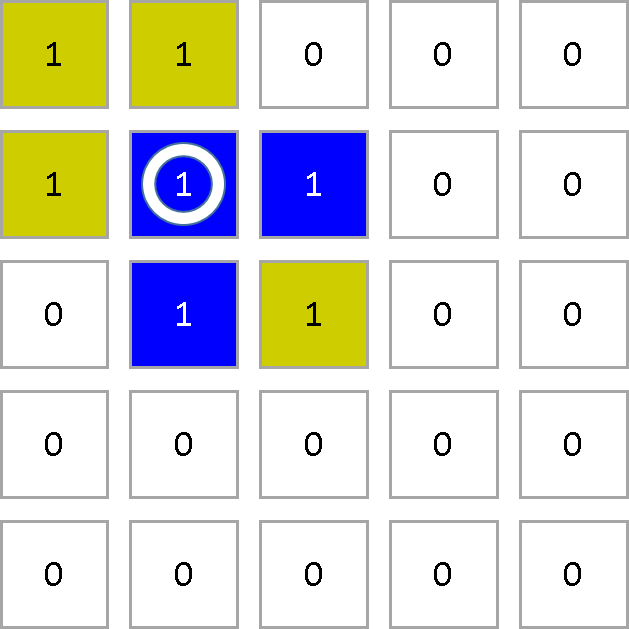}
(c) \includegraphics[width=0.15\textwidth]{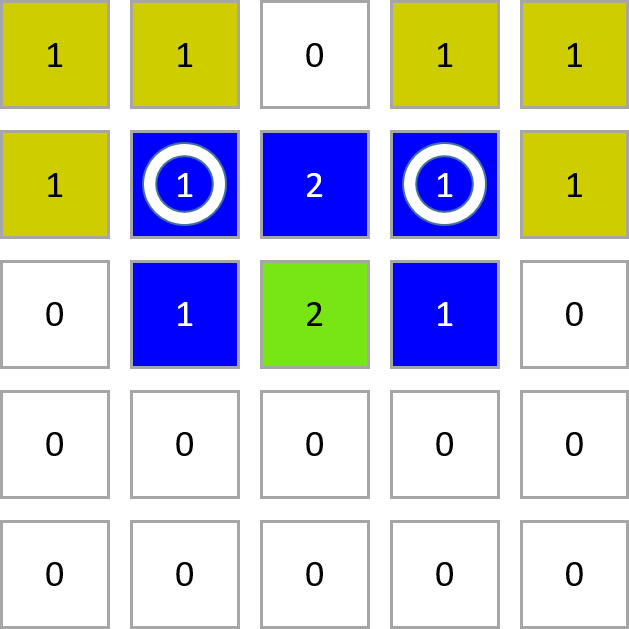}
(d) \includegraphics[width=0.15\textwidth]{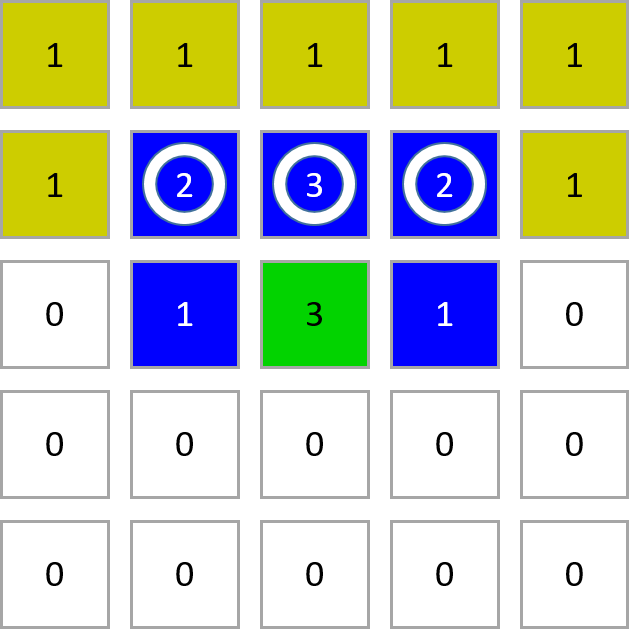}
(e) \includegraphics[width=0.15\textwidth]{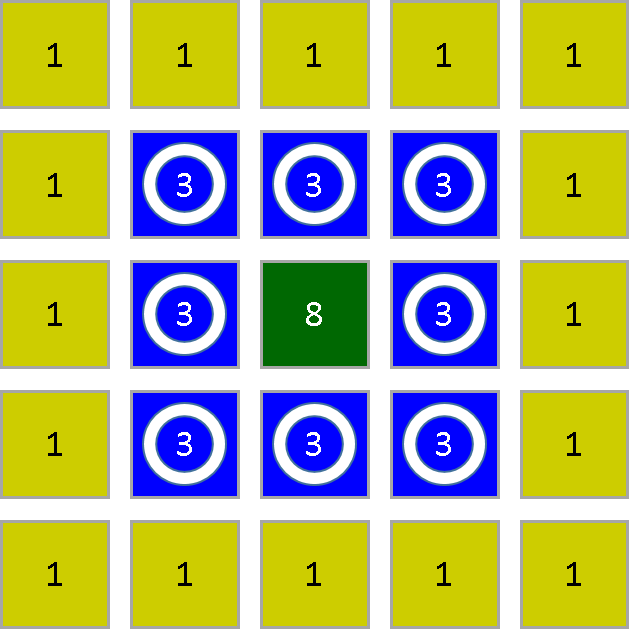}

\vspace{8pt} 
(II)

(a) \includegraphics[width=0.15\textwidth]{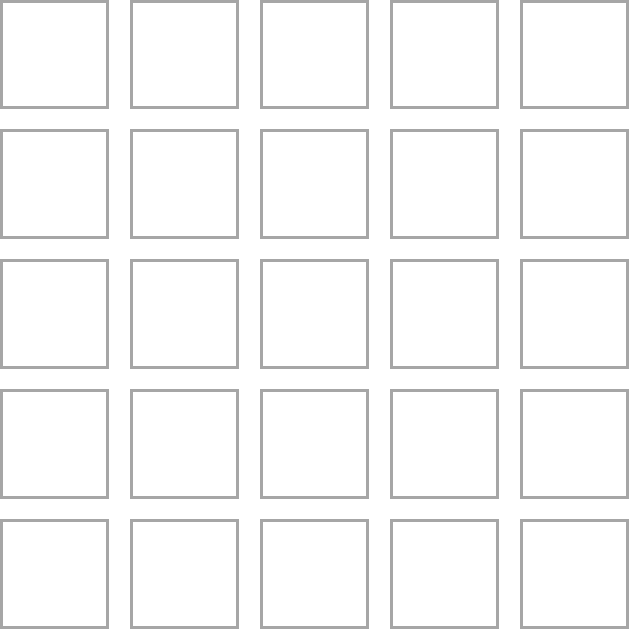}   %$\Rightarrow$
(b) \includegraphics[width=0.15\textwidth]{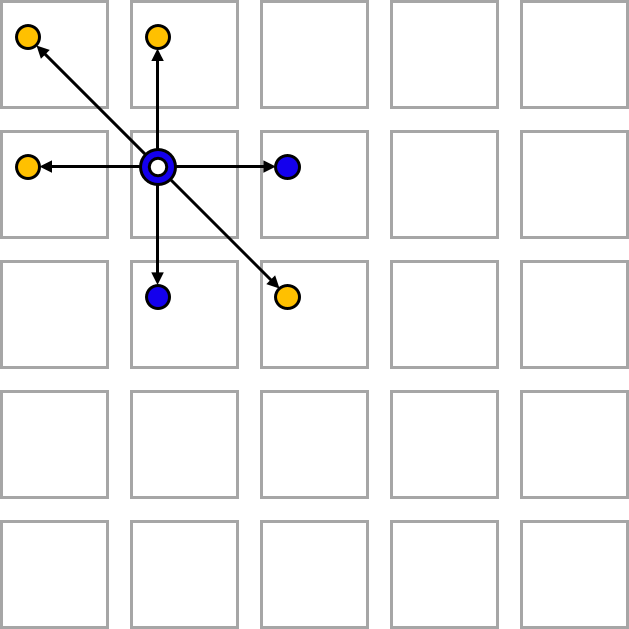}
(c) \includegraphics[width=0.15\textwidth]{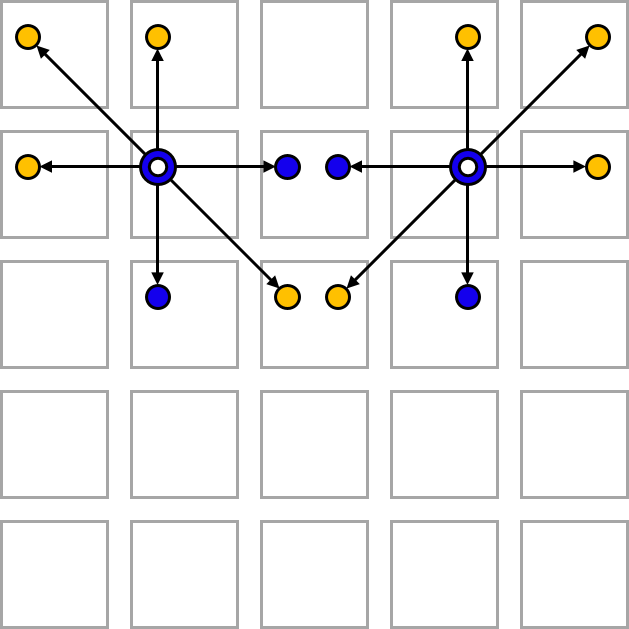}
(d) \includegraphics[width=0.15\textwidth]{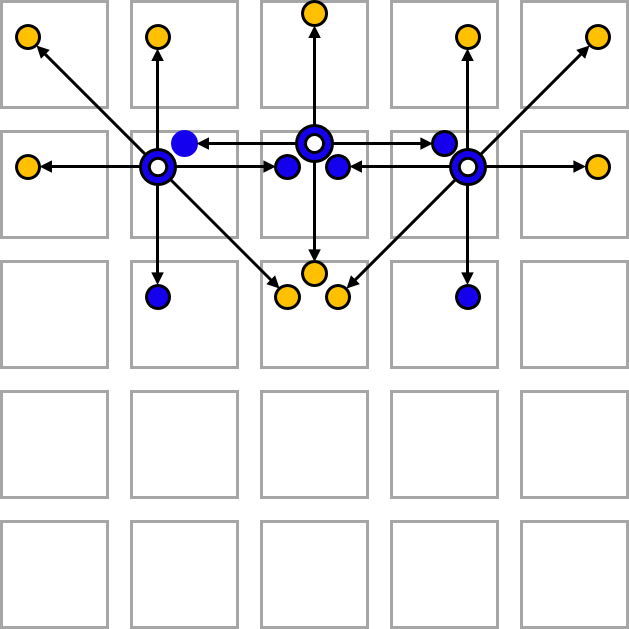}
(e) \includegraphics[width=0.15\textwidth]{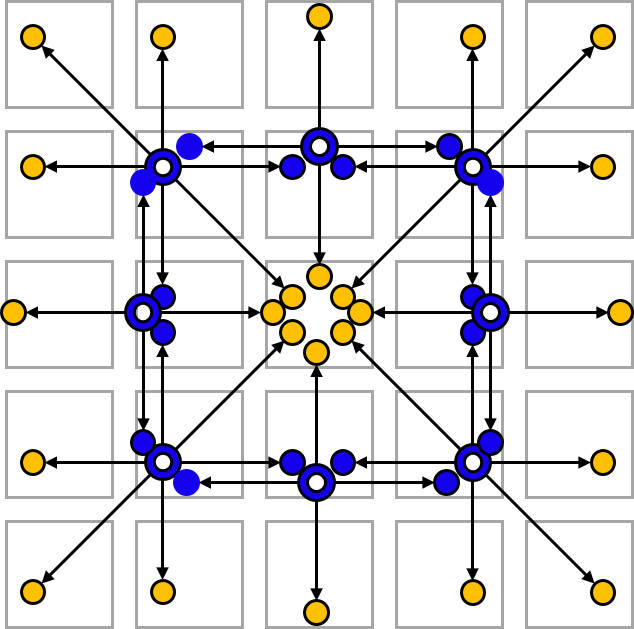}

\caption{%+++Figure GraphTilingLast+++
(I) A $2\times3$ square loop is build by overlapping tiles.
(b) the left upper corner A0 is placed, then
(c) the right upper corner A1, then
(d) a horizontal connection segment B0, and then
(e) B1, A2, B0, A2, B1.
(II) The same problem condidered as a graph covering and coloring problem. 
}
\label{GraphTilingLast}
\end{figure}   
\unskip 
\vspace{11pt}
%===================================================================================== GraphTilingLast

%\begin{bf}
%\color{blue}
Why is $v=3$ the right condition?
Let us consider a horizontal path of one-cells enclosed by zeroes (the hull) that we aim at.
%(We may think of a pipe that encloses some material).
It is possible to form a horizontal straight line of one-pixels using only the line tiles 
B0 =
\texttt{
$\substack{-~0~-
         \\1~1~1
         \\-~0~-}$~~. 
} %%%%%%%%%%%%%%%%%%%%%%%%%%%%%%%%%%%%%%%%%%%%%
\vspace{3pt}

If we place them uncoupled side by side without overlapping ($v=0)$, we yield

\vspace{3pt} 
\begin{center}
\texttt{%%%%%%%%%%%%%%%%%%%%%%%%%%%%%%%%%
$\substack{-~0~-
         \\1~1~1
         \\-~0~-}$  
\textcolor[rgb]{1,0,0}{        
\hspace{-8pt}              
$\substack{-~0~-
         \\1~1~1
         \\-~0~-}$}
\hspace{-8pt}          
$\substack{-~0~-
         \\1~1~1
         \\-~0~-}$   
\textcolor[rgb]{1,0,0}{        
\hspace{-8pt}              
$\substack{-~0~-
         \\1~1~1
         \\-~0~-}$}
\hspace{-8pt}          
$\substack{-~0~-
         \\1~1~1
         \\-~0~-}$   
       $\ldots$ 
}%%%%%%%%%%%%%%%%%%%%%%%%%%%%%%%%%%%%%%%%%%
\end{center}
\vspace{3pt} 

The problem is that 
the line is not totally enclosed by zeroes, therefore the later
described CA rule could try to place ones at the uncovered sites and thereby  the path can be destroyed. 
This means we have to find a way to avoid such situations. % (possible ``holes'' '-' in the pipe). 
If we force the tiles to overlap by one shift count, we yield

\vspace{3pt} 
\begin{center}
\texttt{
%%%%%%%%%%%%%%%%%%%%%%%%%%%%%%%%%%%%%%%%%%%%%\texttt{}
$\substack{-~0~-
         \\1~1~1
         \\-~0~-}$  
\textcolor[rgb]{1,0,0}{          
\hspace{-16pt}            % -5    (-6 or -7)  
$\substack{-~0~-
         \\1~1~1
         \\-~0~-}$}
\hspace{-16pt}          
$\substack{-~0~-
         \\1~1~1
         \\-~0~-}$ 
\textcolor[rgb]{1,0,0}{          
\hspace{-16pt}            % -5    (-6 or -7)  
$\substack{-~0~-
         \\1~1~1
         \\-~0~-}$}
\hspace{-16pt}          
$\substack{-~0~-
         \\1~1~1
         \\-~0~-}$ 
$\ldots$ 
}%%%%%%%%%%%%%%%%%%%%%%%%%%%%%%%%%\texttt{}
\end{center}
\vspace{3pt} 
      
Now the cover level for the ones is either $v=1$ or $v=2$.
The problem is that the hull still contains some '-'. 
The next solution is to choose the maximal overlap $v=3$ by shifting the tiles further together as close as possible. (Note that we do not allow the same tile to overlap with itself). We yield

\vspace{3pt}
\begin{center} 
 %%%%%%%%%%%%%%%%%%%%%%%%%%%%%%%%%%%%%%
\texttt{
$\substack{-~0~~
         \\1~1~1
         \\-~0~~}$  
\textcolor[rgb]{1,0,0}{          
\hspace{-26pt}  
$\substack{~~0~~
         \\1~1~1
         \\~~0~~}$}
\hspace{-26pt}         
$\substack{~~0~~
         \\1~1~1
         \\~~0~~}$ 
\textcolor[rgb]{1,0,0}{          
\hspace{-26pt} 
$\substack{~~0~~
         \\1~1~1
         \\~~0~~}$}
\hspace{-26pt}         
$\substack{~~0~-
         \\1~1~1
         \\~~0~-}$ 
       $\ldots$ 
}%%%%%%%%%%%%%%%%%%%%%%%%%%%%%%%%%\texttt{} 
\end{center}    
\vspace{3pt}      

Now the line is fully enclosed by zeroes, except for the end points which
need a cyclic connection. 
This explanation can easily be generalized if also the other tiles (vertical line, corners) 
are used, as demonstrated for the $3 \times 3$ loop in Fig.~\ref{GraphTilingLast}.
%\color{black}
%\end{bf}
        
\vspace{6pt}
\textit{{Remark.}}
We can consider pattern formation by overlaying tiles as a graph covering and coloring problem. 
Given is a grid graph and a set of sub-grid-graphs (corresponding to the tiles) embeddable in the grid.   
The task is to cover all vertices of the given grid and color the vertices in a way that loops appear.
This approach is equivalent and is illustrated in 
Fig.~\ref{GraphTilingLast} (II) with the grid graph tiles of
Fig.~\ref{TilesLoopUnconnected} (II).
We may call a sub-grid-graph ``\textit{graph tile}''.
Thus we can see the problem as a vertex cover problem  %% todo more/other refs etc.
\cite{1994-vertex-cover1} 
with an additional constraint 
requiring a certain coloring
\cite{2011-vertex-coloring1}.

%%%%%%%%%%%%%%%%%%%%%%%%%%%%%%%%%%%%%%%%%%%%%%%%%%%%%%
\subsection{Templates derived from the tiles}
\label{Templates Derived From the Tiles}
%%%%%%%%%%%%%%%%%%%%%%%%%%%%%%%%%%%%%%%%%%%%%%%%%%%%%%

One of the key ideas is to use templates (local matching patterns)
for testing them everywhere (at any point $(x,y)$) in the CA space.
If we find a template match we call it a \textit{{hit}}.
A hit means that the cell is covered by a pixel of a tile.
We are searching for stable patterns where we have at least one hit~everywhere.

A \textit{template} $L_i$ is a shifted tile.
Let us assume that templates have the size of $m \times m$ pixels. % where $m=2R+1$. 
We have chosen $m$ to be odd because then it is easier to define the unique center  
as the origin for the local coordinates and to handle symmetric templates.
But this choice is not obligatory as long as all templates fit into
the boundaries of $m \times m$.
We index tile and template pixels relative to their center located at
$(x,y)=(0,0)$.
We declare  x-coordinates to increase rightwards and y-coordinates~downwards.

For a given tile $T$ (the \textit{{generating}} tile), we derive a set of templates by shifting the 
generating tile vertically and/or horizontally.
We consider only the valid pixels with a value of 0/1 for the process of deriving the templates.
We call a valid pixel \textit{{reference pixel}} when it is selected for deriving a specific template.
For each reference pixel with the (relative) coordinates $(x_{ref(i)}, y_{ref(i)})$,
we shift the tile in such a way that after shifting, the reference pixel 
occupies/arrives at the center $(0,0)$ of the derived template, i.e.,

\begin{itemize}
\item 

for each reference point $i$ at $(x_{ref(i)}, y_{ref(i)}):$
$L_i = \textit{shift}(-x_{\textit{ref}(i)} , -y_{\textit{ref}(i)}, T)$ %for ~each ~reference ~point (x_{ref(i)}, y_{ref(i)})$

where the operator  $\textit{shift}(\Delta x, \Delta y, T )$ shifts a tile $T$ by the given offsets, and~is
\newline\indent
shifting-in null pixels and deleting shifted-out pixels. 
\end{itemize}

%===================================================================================== Templates
\vspace{-3pt}
\BEGINFIGURE          
\includegraphics[width=0.8\textwidth]{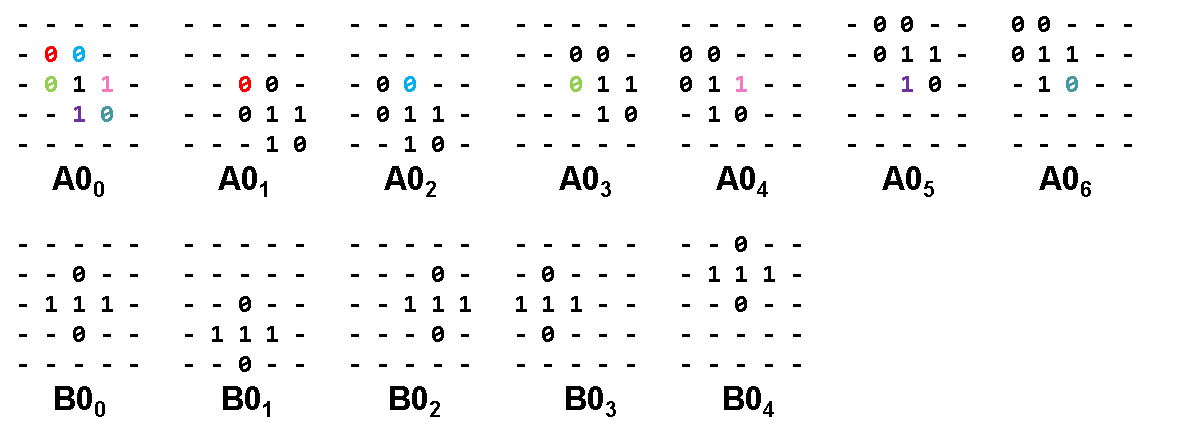} %12cm
\caption{
The templates derived from the corner tile $A0 = A0_0$
and the horizontal  line tile  $B0 = B0_0$.
Templates are shifted tiles. 
}
\label{Templates}
\end{figure}   
\unskip 
\vspace{11pt}
%===================================================================================== Templates            
Figure~\ref{Templates} shows the templates derived from $A0$ as defined in Figure~\ref{TilesLoopUnconnected}.
$A0 = A0_0$ is the generating tile and also the first (the main) template
that needs no shift because it contains a valid pixel in its center already (what we assumed). 
Now, we consider the  pixel  ${A0}{[-1,-1]}$, in red.
A0 is shifted by $(1, 1)$ steps, (1 rightwards, 1 downwards), which yields $A0_1$.
$A0_2$  is A0 shifted by $(0, 1)$, where the red pixel occupies the center. 
$A0_3$  is A0 shifted by $(1, 0)$, and~so on. 
As a result, we obtain seven templates derived from $A0$.
In a similar way, we obtain the templates from the remaining tiles.
We get from the 4 corner tiles with 7 valid reference pixels $4 \times 7$ = 28 templates, and
from the 2 line tiles with 5 reference pixels $2 \times 5$ = 10 templates $L_i$, altogether 38. 
We may reduce the number of templates by joining them using classical methods of
minimization logical functions where  the center value is the output and the 
neighboring pixels are the inputs.  
The templates (to fit in a square array) are larger than the given tile because of shifting,
the radius of the templates is twice the radius of the given tile.

\textit{{Remark.}} In general, we can define a template set independently of certain tiles, but~
this is a general topic that will not be followed here.

%%%%%%%%%%%%%%%%%%%%%%%%%%%%%%%%%%%%%%%%%%%%%%%%%%%%%%
\subsection{The rule in detail}
\label{The Rule Details}
%%%%%%%%%%%%%%%%%%%%%%%%%%%%%%%%%%%%%%%%%%%%%%%%%%%%%%
In the next Sect. 
\ref{RuleGeneralForm}
the designed CA rule is explained in detail.
The rule is somehow general and looks quite complex.
It includes several optional conditions which can separately be enabled.
Depending on the enabled conditions 
%specialized 
%rule variants  (\textit{Rule0..4}) are
rule variants  (Rule0 .. Rule4) are
derived and used when appropriate. These specialized rules 
%(rule variants) 
are described in Sect.
\ref{The Rule Variants}.
%The following CA rule summarizes  different more simple rule variants.
The usage and motivation for the rule variants becomes clearer when
reading Sect.~\ref{Simulations}.

%%%%%%%%%%%%%%%%%%%%%%%%%%%%%%%%%%%%%%%%%%%%%%%%%%%%%%
\subsubsection{The CA rule in a general form}
\label{RuleGeneralForm}
%%%%%%%%%%%%%%%%%%%%%%%%%%%%%%%%%%%%%%%%%%%%%%%%%%%%%%
The following CA rule 
is able to evolve loop patterns
and it is inserted into the simulation algorithm (Fig.~\ref{simalgorithm}).
The main idea is 
to inject noise if no template hit is recognized
in order to drive the evolution to a stable pattern.
In such a pattern, there is everywhere a template hit, i.e.,
every CA cell is either equal to the center of a tile A
or it is equal to a pixel of a tile B, and
where that pixel is the center of a template derived from B.

The rule consist of three computational steps in sequence. 
(1. TEST) 
All templates $A_i$ are tested against the current CA neighborhood at the
selected site $(x,y)$ where the template center is excluded from the test.
The number of hits $h_0$ for $(s=0)$ and $h_1$ for $(s=1)$ are stored.
(2. ADJUSTMENT)
If there is a template hit, the center of the template defines the new state.
This operation is called \textit{adjustment}:
either the CA neighborhood corresponds fully to the template, or only the center is false
and then it is corrected.
In the case of several hits, the first hitting template is used for adjustment.
(3. NOISE INJECTION)
Noise is injected under certain conditions with certain probabilities. 
The main necessary condition is the ``\textit{path condition}'' saying that
a path cell has to show three one hits ($h_1=3$).
Other ``noise injection conditions'' are explained in the following details of the 
three steps.

%________________________________________________________________
\begin{enumerate}
	\item (TEST)

$(h_0, h_1, k) \leftarrow \textit{ComputeHits of all Templates} ~L_i \in\{A_{0} \ldots A_5, B_0, B_1 \}$

\vspace{5pt}
where 

\leftskip4mm

$h_0$ = number of zero-hits (hits with template center value = 0),

$h_1$ = number of one-hits (hits with template center value = 1),

$k$ = index of the first  template match $L_k$.

%\begin{addmargin}[1cm]{2cm}\end{addmargin}

\leftskip0mm
\vspace{5pt}
If there are several hits, the first hit is used for adjustment in the second step.
The templates are tested in a fixed random order for the purpose of avoiding any preference. 
It would be possible to use a certain order as to give certain tiles a higher priority. 
To which extent the order influences the pattern structure is a topic of further research.\\
%\leftskip0mm
%________________________________________________________________

%________________________________________________________________

	\item   (ADJUST)

\[
s' =
\left
\{
\begin{array}{llll} 
   
          \textit{center}(L_k)  	& \textit{if}   &h>0       &~~~(a)\\     %  ~at~ x,y)    

                             s    & \textit{else} &          &~~~(b)\\                     
                                                                                  
\end{array}
\right.
\]

where 

\leftskip4mm
  $h=h_0+h_1$,

  $\textit{center}(L_{k})$
  is the value of the center pixel of the  match of template $L_k$. 

  \vspace{5pt}
  In fact there are two cases: (i) the center pixel is already  equal to the state s of the cell,
  then the template is completely recognized and nothing needs to be done ($s'=s$),
  or
  (ii) only the center pixel of the template is not correct, then it is adjusted to the correct value.\\
%\leftskip0mm
%________________________________________________________________

%_______________________________________________________________
	\item  (NOISE INJECTION)
  
\[
s \leftarrow %s'' =
\left
\{
\begin{array}{llll} 
   
	               R_P & \textit{if}   &(s'=1) \wedge P     &~~~~(a)\\
								 R_Q & \textit{if}   &(s'=0) \wedge Q     &~~~~(b)\\     %  ~at~ x,y)   

                s'  & \textit{otherwise} &                 &~~~~(c)\\                     
                                                                                  
\end{array}
\right.
\]

where 

  \leftskip4mm
  
   %$P= P_1'  \vee  P_2'$
   %$P= P_1^\delta  \vee  P_2^\delta$
   $P= \hat{P_1} \vee  \hat{P_2}$

    \leftskip8mm

    $\hat{P_1} = P_1  \wedge  Decision(\pi_{P_1})$  \hfill-- used in all rule variants \textit{Rule 0 .. 4}
    
       \leftskip12mm	
        $P_1=(h_1 \neq 3)$            
    
    \leftskip8mm

    $\hat{P_2} = P_2 \wedge Decision(\pi_{P_2})$  \hfill -- used in \textit{Rule 3,4}
    
        \leftskip12mm	
        $P_2 = \textit{TRUE} \textit{~if exist two consecutive~} (h_0=0)-\textit{cells at distance} ~2 \textit{~and} ~3$

  \vspace{5pt}
  \leftskip4mm  
  $Q= \hat{Q_0} \vee {Q_1}$

    \leftskip8mm
		 %$\hat{Q_0}= Decision(\pi_{Q_0}) \wedge Q_0$
		$\hat{Q_0}=  Q_0 \wedge Decision(\pi_{Q_0})$
    
		    \leftskip12mm		
         $Q_0= (h_0 = 0)  \wedge NOT(E) $    \hfill   -- $(h_0 = 0)$ used in \textit{Rule 1,2,3,4}

         $E=(E_2  \wedge  E_3) \vee Decision(\pi_{E})$   \hfill   -- used in \textit{Rule 2,3,4}
				
         \leftskip12mm			
			  $E_{i=2,3} =  \textit{Exists a cell at distance i (orthogonal or diagonal) with~} h_1=3$

    \leftskip8mm  
    $Q_1=  (h_0 = 8)   \wedge   Decision(\pi_{Q_1})$  \hfill -- used in \textit{Rule 4} only    
      
\vspace{5pt}
\leftskip0mm
\end{enumerate}
%_______________________________________________________________

%_______________________________________________________________
\begin{itemize}
	\item 

$R_P$ and $R_Q$ are values that drive the evolution by setting them to a value unequal to the current state
with a certain probability.
Here we are using $R_P=R_Q=R$ randomly chosen as either 0 or 1.
In this way noise is injected if the condition $P$ becomes true for a one-cell, 
or if $Q$ becomes true for a zero-cell.
  \vspace{5pt}

	\item Conditions for one-cells (with state $s'=1$):
  
    %_____________________________________________________
    \begin{itemize}
    \item 
		\NEW{
	  The condition $P_1$ is the main condition  
    that realizes the \textit{loop path condition}.    
    It means that three one-pixels (of three tiles in sequence) shall overlap at the current site, and one of them is there centered.
		}		
    No more noise will be injected if there are exactly three $h_1$-hits. 
    Thereby occurring path cells are stabilized. 
    
      \begin{itemize} %[leftmargin=5.5mm,labelsep=2.5mm]
      \item 
       $Decision(\pi)$ is a function that is TRUE if a trial is successful under probability $\pi$,
       otherwise FALSE.
      The condition following it can be switched OFF\-/ON by setting $\pi=0/1$.
      In the language Python for instance, this function can be defined as: 
      ``def decision(probability): return random() < probability'' 
      \end{itemize}    
      
        \item
    The optional condition $P_2$ forbids a path cell to establish if there 
    are uncovered cells (with $h_0=0$) at a distance 2 AND 3 in orthogonal or diagonal direction. 
    The purpose is that thereby a path can move to the direction of uncovered areas.

    \end{itemize}
    %_____________________________________________________

    \item Conditions for zero-cells (with state $s'=0$):
    %_____________________________________________________
    \begin{itemize}
      \item 
      
      If the condition $(h_0=0)$ becomes true then 
      the cell is uncovered and noise is injected.
      The main intention is to include such cells into the pattern evolution.   
      If it is not used, then uncovered areas may survive.
      
			\item
			The  optional condition $E$ defines  an exception which allows certain  situations to be stable by 
			suppressing noise injection.
      Condition $E_{2/3}$ allows an uncovered zero-cell to be stable if it is situated at a distance of 2/3			    orthogonally or diagonally from a path cell with $h_1=3$.  
        The idea is to allow a small amount of uncovered cells in a pattern in order  to 
        allow more possible patterns to be stable. 
        
      \item            
      This optional condition $Q_1$ can be used to forbid certain configurations with a local overlap level
      of $H$, thereby giving preference to certain loop patterns. 
      This is a very special condition, and it exemplifies how the rule can be modified further.

    \end{itemize}
    %_____________________________________________________          

\end{itemize}
%________________________________________________________________

%%%%%%%%%%%%%%%%%%%%%%%%%%%%%%%%%%%%%%%%%%%%%%%%%%%%%%
\subsubsection{The rule variants}
\label{The Rule Variants}
%%%%%%%%%%%%%%%%%%%%%%%%%%%%%%%%%%%%%%%%%%%%%%%%%%%%%%

The behavior of the rule depends on the 
activated conditions $C_i\in \{P_1, {P_2}, Q_0, Q_1, E\}$.
We write  \textit{RuleK} = $Rule(C_1/C_2/...)$ if the conditions $(C_1, C_2, ...)$
are active. 
The probabilities 
$\pi_{P_1},\pi_{P_2}, \pi_{Q_0},\pi_{E},\pi_{Q_1}$ 
can also be used to disable 
an associated conditions if $\pi_i=0$. 
For an active condition \textit{Decision}  $\pi_i=0.5$ was used if not specified otherwise.
Other values can influence the speed of convergence and they can give preference to 
certain loop structures (for instance with a small  amount of uncovered cells on average),
 but this issue was not further investigated. 

Depending on the activated conditions we will use five rule variants 
that are special cases of the general rule. 

\begin{itemize}
	\item 
 
	 $Rule(P_1)$ = \textit{Rule0}, the 'PATH-ONLY' rule.

  It injects noise to a one-cell if $(h_1 \neq 3)$.  
	Only the path condition $P_1$ is enabled.
  As the path condition is a necessary condition it is activated in all following rules
  by default.

	\item
	$Rule(P_1/Q_0)$ = \textit{Rule1}, the 'UNCOVERED-ZERO' rule.
  
 % It injects additional \textit{Noise if} $(s'=0)\vee(h_0 =0)$. 
   It injects  noise to a zero-cell if $(h_0 =0)$. 
  The purpose is to destroy uncovered cells / areas.
In addition to the path condition the condition $Q_0$	 (noise injection for uncovered zero-cells) is active. 
%The other conditions ($Q_1, E)$ are disabled by setting $\pi_{Q_1}=0,\pi_{E}=0$.

\item
$Rule(P_1/Q_0/E)$ = \textit{Rule2} , the 'UNCOVERED-EXCEPTION' rule.

No noise is injected for a uncovered zero-cell if it sees a path cell at a distance 2 or 3.
Thereby previously uncovered toggling zero-cells at distance 2 and 3 can stabilize.
The noise condition of the previous \textit{Rule 1} is weakened (no noise is injected) if the exception $E$ becomes true. 
In addition to  the \textit{Rule1}-- conditions	
the condition $E$ is active (both conditions $E_2$ and $E_3$ are active).

\item
	$Rule(P_1/P_2/Q_0/E)$ = \textit{Rule3}, the 'PATH-DESTROY' rule.
  
Noise is injected for a one-cell if it sees uncovered zero-cells at a distance of 2 AND 3.
%if tries to settle near such an area (uncovered at distance 2 AND 3).  
The purpose is to break transient loops / paths which are already are stable (or partially stable) and try to settle near large uncovered areas. 	
In addition to the \textit{Rule 2} -- conditions the condition $\hat{P_2}$ is active.

	\item
	$Rule(P_1/P_2/Q_0/E/Q_1)$ = \textit{Rule4}, the 'FORBID3x3' rule.

In addition to the \textit{Rule3}-conditions the condition $Q_1$	is active in order to drive the evolution to
a certain subset of all possible patterns. 
As an example we want to forbid (by noise injection) the cover level $v=8$ for zero-cells by setting $H=8$.
Thereby small $3\times 3$ loops are destroyed and will no longer appear in the final stable pattern.
\end{itemize}

In the next section we will use these five rule variants in the simulation, and we will learn 
about their abilities.
Note that we can define more variants depending on the combination of enabled noise injection conditions. 
We could use for instance 	$Rule(P_1/P_2)$ with different probabilities $\pi_{P_1},\pi_{P_2}$ which
can produce good loop patterns, too.

%%%%%%%%%%%%%%%%%%%%%%%%%%%%%%%%%%%%%%%%%%%%%%%%%%%%%%
\section{Simulations}  \label{Simulations}
\label{S4}
%%%%%%%%%%%%%%%%%%%%%%%%%%%%%%%%%%%%%%%%%%%%%%%%%%%%%%

We increase the field size step by step from $3\times 3$ to $7\times 7$
and at the same time we increase the complexity of the rule variant (from
\textit{Rule0} to \textit{Rule4}).
In this way %               it is intended to show and motivate 
(i) the rule variants could be (and more or less were) developed, and 
(ii) we learn about their properties.
%we argue what for the rule variants are good and needed. 

We discuss the rules' properties depending on the field size: 
\textit{Rule0} and \textit{Rule1} for $n=3$,
\textit{Rule1} for  $n=4$,
\textit{Rule0 .. 2} for  $n=5,6$, and 
\textit{Rule0 .. 4} for $n=7$.
Closed space filling curves (SPC) can appear for $n=3,7,11,(+4) ..$.
Therefore we then considered $n=11$ with \textit{Rule4} and found some SFC among many other loops.
%Unfortunately the appearance  of SPC for higher $n$ is rare (because of the strong symmetries).
Then we show loop patterns of size $32\time32$ for illustration. 

\newpage
%%%%%%%%%%%%%%%%%%%%%%%%%%%%%%%%%%%%%%%%%%%%%%%%%%%%%%
\subsection{Pattern of size $3\times3$}
%%%%%%%%%%%%%%%%%%%%%%%%%%%%%%%%%%%%%%%%%%%%%%%%%%%%%%

Initially we start the simulation with a random configuration if not specified differently. 
First we tried a very simple rule (\textit{Rule0}) where only the path condition $P_1$ is used, 
all other conditions ($P_2, Q)$ are FALSE.
In the simulation we reach two possible fixed points, 
the $3\times3$ square (Fig.~\ref{u4x4-1}a,b) or  the all-zero configuration (Fig.~\ref{u4x4-1}c,d).
%%%%%%%%%%%%%%%%%%%%%%%%%%%%%%%%%%%%%%%%% u4x4-1
\BEGINFIGURE
%\sidecaption            %%max 7.8cm with sidecaption
(a)\includegraphics[width=2cm,height=2cm]{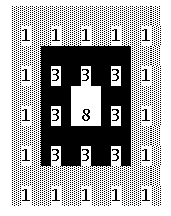}  %width=0.2\textwidth, 
(b)~\includegraphics[width=1.9cm,height=1.9cm]{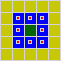}
~~~~(c)~\includegraphics[width=2cm,height=2cm]{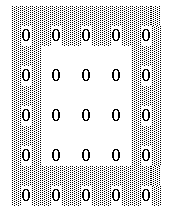}  %width=0.2\textwidth, 
(d)~\includegraphics[width=1.9cm,height=1.9cm]{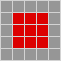}
\caption{%++u4x4-1;++
For $n=3$ there appears only one  loop, the $3 \times 3$ square.
(a) Shows the simulation result in black/white where the numbers represent the cover level $v$. 
(b) The result (a) in color representation, with the tile centers marked by red dots.
(c, d) Special fixed point: the all-zero configuration. Uncovered cells with $v=0$ are colored in red. 
}
\label{u4x4-1}   
\end{figure}
%%%%%%%%%%%%%%%%%%%%%%%%%%%%%%%%%%%%%%%%% u4x4-1

Usually we like to avoid the all-zero fixed point.
The problem is solved when we specify a conditions $Q$ for noise injection if the cell state $s'=0$.
The simplest way (\textit{Rule1}) is to use the condition $Q_0$ which raises noise if a zero-cell shows no hit.
A hit are strongly related to the cover level $v$ because it converges to it when the pattern becomes stable. 
So when we activate this condition, the problem is solved (at least for this field size), and the all-zero configurations appears not any more.

\NEW{
Fig.~\ref{simsequence3x3} illustrates  how a $3\times3$ square loop evolves stepwise in micro-steps
starting with an all-one configuration (for all the internal 9 cells). 
}
%%%%%%%%%%%%%%%%%%%%%%%%%%%%%%%%%%%%%%%%% simsequence3x3
\BEGINFIGURE
%\sidecaption            %%max 7.8cm with sidecaption
\includegraphics[width=0.9\textwidth,height=3.8cm]{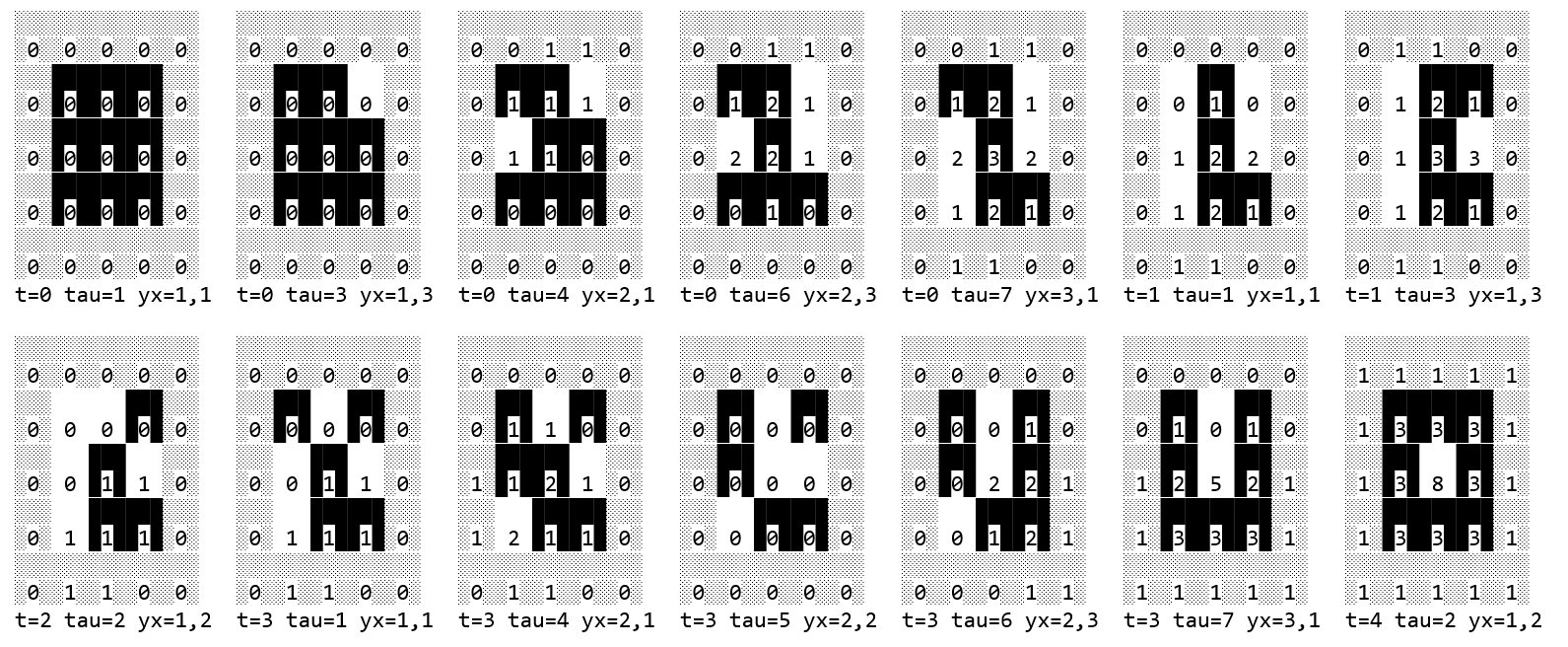}  %ok aber geringe qualität
\caption{%++simsequence3x3++ 
A simulation sequence in steps ($t$) and micro-steps (\textit{tau}). 
In order to follow the state changes more easily, the cells were selected here in sequential order.
Only patterns that have changed are shown. 
The numbers in the cells give the cover level $v$.
\NEW{
Initially all the inner active cells were set to one, and the border cells are fixed to zero.}
}
\label{simsequence3x3}   
\end{figure}
%%%%%%%%%%%%%%%%%%%%%%%%%%%%%%%%%%%%%%%%% simsequence3x3

%%%%%%%%%%%%%%%%%%%%%%%%%%%%%%%%%%%%%%%%%%%%%%%%%%%%%%
\subsection{Patterns of size $4\times4$}
%%%%%%%%%%%%%%%%%%%%%%%%%%%%%%%%%%%%%%%%%%%%%%%%%%%%%%

For a field of size $4\times4$ there are only 4 possible loops (Fig.~\ref{4x4pattern})
which can easily be constructed.
In the simulation the \textit{Rule1}
was applied.

%%%%%%%%%%%%%%%%%%%%%%%%%%%%%%%%%%%%%%%%% 4x4pattern
\BEGINFIGURE
\includegraphics[width=0.6\textwidth]{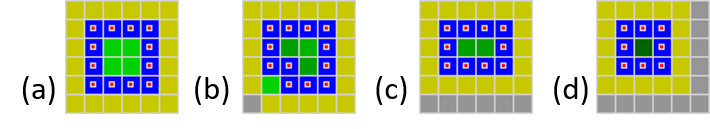}  %ok aber geringe qualität
\caption{%++4x4pattern++ 
The four possible loops of a field of size $4\times 4$. The length of the loops are $(12,12,10,8)$ for (a, b, c, d).}
\label{4x4pattern}   
\end{figure}
%%%%%%%%%%%%%%%%%%%%%%%%%%%%%%%%%%%%%%%%% 4x4pattern

%%%%%%%%%%%%%%%%%%%%%%%%%%%%%%%%%%%%%%%%%% 4x4pattern
%\begin{SCfigure}
%\includegraphics[width=0.6\textwidth]{Figures/4x4pattern.png}  %ok aber geringe qualität
%\caption{%++4x4pattern++ 
%The four possible loops of a field of size $4\times 4$. The length of the loops are $(12,12,10,8)$ for (a, b, c, d).}
%\label{4x4pattern}   
%\end{SCfigure}
%%%%%%%%%%%%%%%%%%%%%%%%%%%%%%%%%%%%%%%%%% 4x4pattern

\newpage
%%%%%%%%%%%%%%%%%%%%%%%%%%%%%%%%%%%%%%%%%%%%%%%%%%%%%%
\subsection{Patterns of size $5\times5$}
%%%%%%%%%%%%%%%%%%%%%%%%%%%%%%%%%%%%%%%%%%%%%%%%%%%%%%

For a field of size $5\times5$ we found  33 possible loop patterns (Fig.~\ref{ALL5x5}),
%
%%%%%%%%%%%%%%%%%%%%%%%%%%%%%%%%%%%%%%%%% ALL5x5
\BEGINFIGURE
%\sidecaption            %%max 7.8cm with sidecaption

\includegraphics[width=0.9\textwidth]{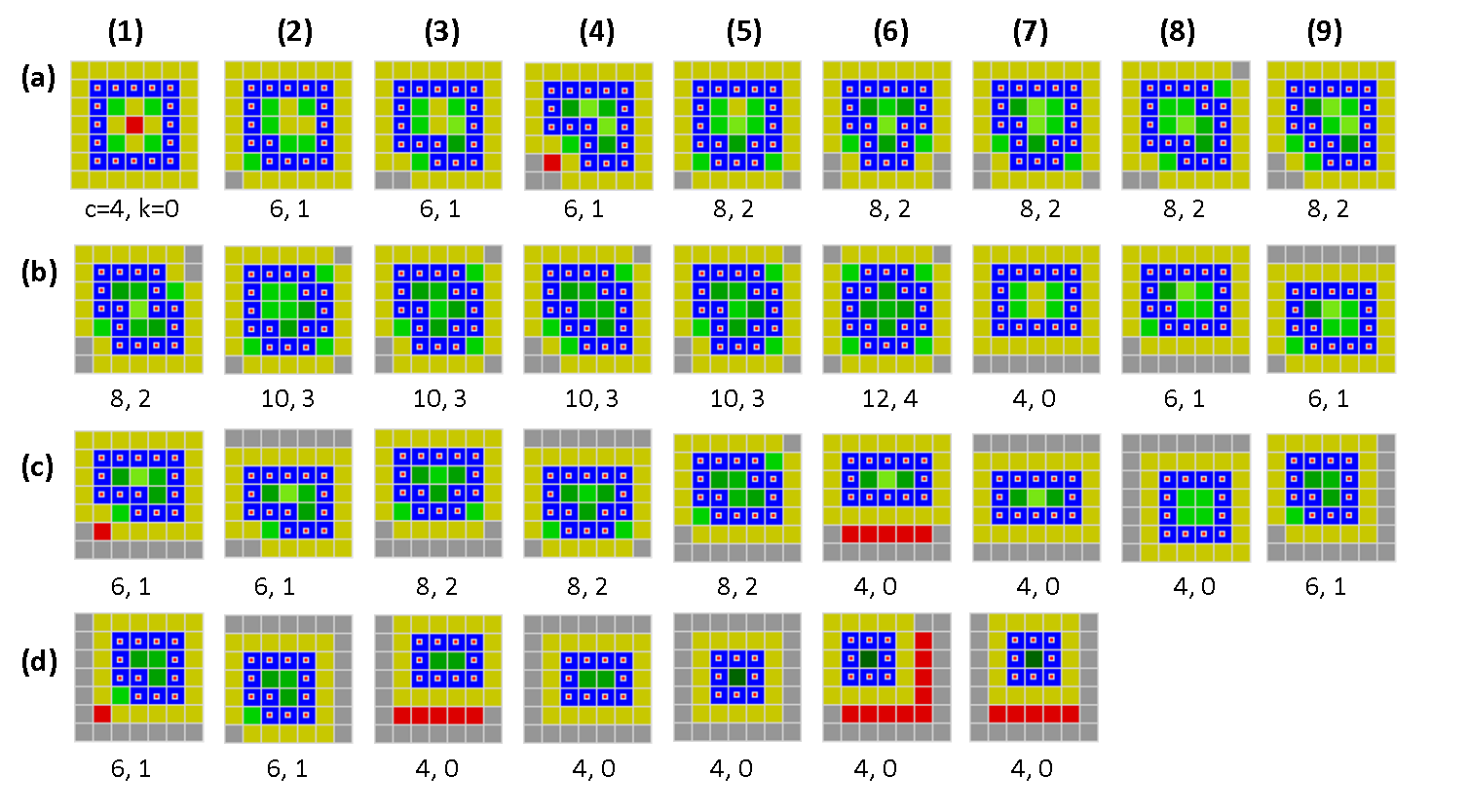}  %ok aber geringe qualität
\caption{%++ALL5x5++ 
All the detected loop patterns of size $5\times 5$. 
Uncovered cells, which are not on the boundary, are marked in red. The number of corners $c$ and %the length of the loops  $g$ are given.
the number of inner/convex corners $k$ are given.
The following loops are equivalent under shift:  
$(b8, b9), (c1, c2), (c3, c4), (c6, c7), (c9, d1, d2), (d3, d4), (d5, d6, d7)$.
The patterns are ordered with respect to  
	the size of the rectangle  that encloses a loop.} 
	\label{ALL5x5} 
\end{figure}
%%%%%%%%%%%%%%%%%%%%%%%%%%%%%%%%%%%%%%%%% ALL5x5
%
where symmetric patterns under rotation and mirroring are not shown/counted. 
Some of the loops show the same structure but are shifted which results in a different 
covering configuration, often with uncovered cells (marked in red if not on the boundary). 
If patterns with the same loop structure (equivalents under shift) are counted only once
then we obtain only 24 different loops.
Only a subset of all patterns (Fig.~\ref{ALL5x5}) will be evolved depending on the 
applied rule variant:

\begin{itemize}
	\item \textit{Rule0}.  
  All the patterns as shown in  (Fig.~\ref{ALL5x5}) can appear as stable fixed points. 
  The special case 'all-zero' may also appear (not shown).

  \item 
	\textit{Rule1}. %   \textit{Rule1}. 
  Noise is injected for uncovered zero-cells (marked in red). 
  Thereby the patterns c6, d6, d7 are only transients and  will not be stable.
  %%NEW??Thereby the patterns that contain red cells will not be completely stable because the red cells are toggling.
  Let us consider the simulation shown in Fig.~\ref{simRule1-5x5}
  with c6 as initial/transient configuration.
  %
  %
  %%%%%%%%%%%%%%%%%%%%%%%%%%%%%%%%%%%%%%%%% simRule1-5x5
\BEGINFIGURE
%\sidecaption            %%max 7.8cm with sidecaption

\includegraphics[width=0.9\textwidth,height=7cm]{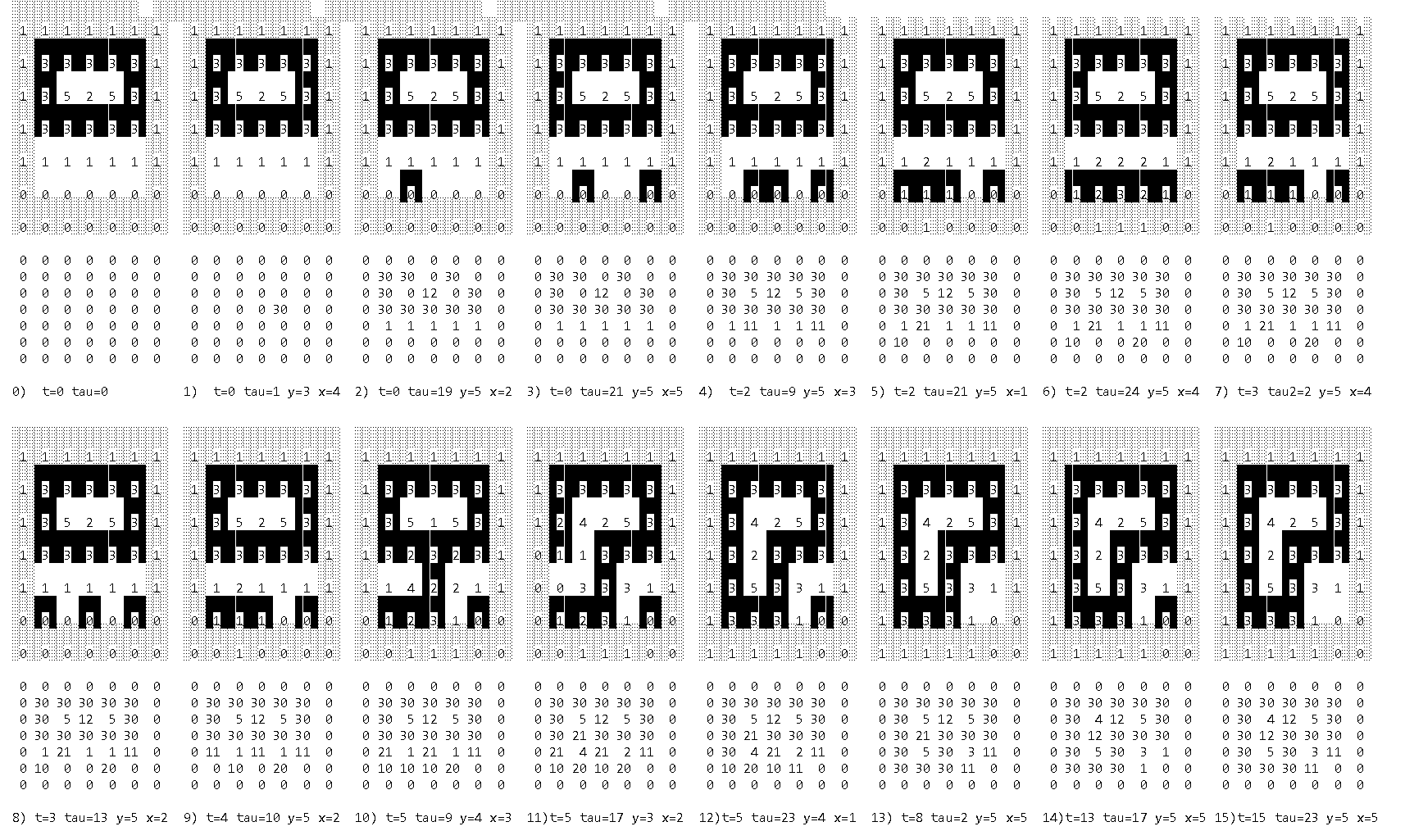} 
\caption{%++simRule1-5x5++ 
A simulation of the pattern 
(Fig.~\ref{ALL5x5} c6) used as 
 initial pattern with snapshots at \textit{tau}  micro-steps when the pattern changed. 
\textit{Rule 1} was applied.
The first and third row show how the pattern evolves (the cell states with their cover
levels), and the second and forth row give the corresponding hit values  as
$10h_1 + h_0$.
}
\label{simRule1-5x5}   
\end{figure}
%%%%%%%%%%%%%%%%%%%%%%%%%%%%%%%%%%%%%%%%% simRule1-5x5
  %
  %
  In the first and third row it is shown how the pattern evolves
  (the cell states with their cover levels), 
  and in the second and forth row the corresponding hit values  are given as $10h_1+h_0$.
  Only patterns that have changed are shown.
  The hit values may have changed in between. 
  Note that the hit value at the current position $(y,x)$ is up-to-date,
  but at other positions they can be false because they correspond to the last update in the past.
  Such 'false' hits in the neighborhood are another source of noise for the evolution
  when conditions $P_2$ or $E$ are activated.
  Fortunately this noise does not destroy totally the evolution to the aimed patterns
  (as a result observed during many experiments, but a formal proof is needed). 
  The hit values $h_{0/1}$ converge to the cover levels for $s=0/1$
  when the pattern stabilizes (see for instance the last pattern (15) and compare the cover levels with the hit values). --   
  In this simulation we can observe the following events:
  \begin{enumerate}
    \item  
    
  An uncovered zero-cell at the lower border at the distance 2 from  a path cell changes to one
  (e.g. sub figures  (2) -- (9)).
  
   \item 
  A template hit $h_1$ for tile B1 appears (directly near the path) for a zero-cell
  between a one-cell at the bottom and a path cell (belonging to the lower part of the loop)
  (e.g. (9)).
  
   \item 
  The former zero-cell changes to one (by adjustment) which looks like a branch from the loop
  (e.g. (10)).
  
   \item 
  The path condition for the involved path cell (near the branching, $y=3, x=2$) is violated,
  and this path cell changes to zero (see (10, 11)). Thereby the loop is broken. 
  
  \item
  There may remain some toggling cells 
  (here the rightmost bottom corner cell, see (12) to (15)).
  
   \end{enumerate}
  
  Using \textit{Rule1} the patterns a1, a4, c1, d1 of 
	 Fig.~\ref{ALL5x5}
	can still appear but the uncovered cells (red marked) are toggling between 0 and 1,
  because an uncovered zero-cell is changed to one, and a one-cell that is not a path cell is changed to zero. 
  The already formed loops in such patterns remain stable. So we may obtain patterns that are partially unstable within the whole space. 
    
  \item \textit{Rule2}.
  Now we want to avoid toggling cells and change them into stable zero-cells.
  Thereby we can evolve all patterns being stable with partially uncovered cells which already were
  reached by the simple \textit{Rule0},
  including (c6, d6, d7) which were excluded by \textit{Rule1}. 
  Now the condition $E=E_2 \vee E_3$ is activated with probability $\pi_{E}=0.5$.
  The condition $E_{2/3}$ allows uncovered zero-cells to survive which stay at a distance of 2/3 (orthogonally or diagonally) apart from a path cell.  
 As the result we aimed at,  all the shown 33 patterns can be reached as fixed points.
  
  \textit{Remark}: Condition $E_3$ needs not to be active  for this field size of $5 \times5$ 
  because loop-patterns with uncovered cells at distance 3 are not possible. 
  $E_3$ was introduced for larger fields where uncovered cells of distance 3 can appear
    which shall be stabilized. 
  
\end{itemize}

%%%%%%%%%%%%%%%%%%%%%%%%%%%%%%%%%%%%%%%%%%%%%%%%%%%%%%
\subsection{Patterns of size $6\times6$}
%%%%%%%%%%%%%%%%%%%%%%%%%%%%%%%%%%%%%%%%%%%%%%%%%%%%%%

Not all possible $6 \times 6$  loop patterns can be displayed here, therefore only a selection 
(Fig.~\ref{ALL6x6}) is presented.
%
%
%%%%%%%%%%%%%%%%%%%%%%%%%%%%%%%%%%%%%%%%% ALL6x6
\BEGINFIGURE
%\sidecaption            %%max 7.8cm with sidecaption

\includegraphics[width=0.9\textwidth]{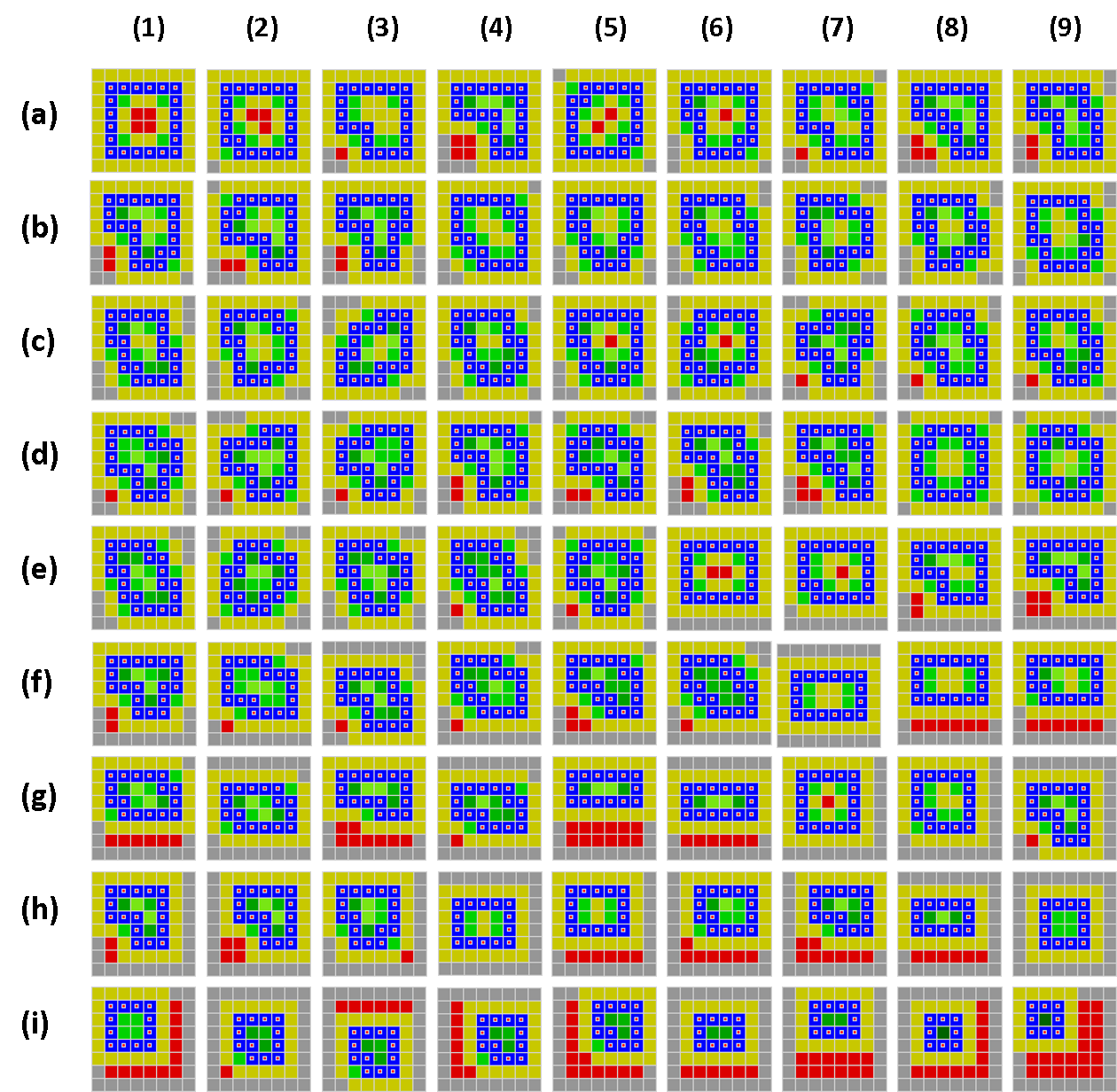}  %ok aber geringe qualität
\caption{%++ALL6x6++ 
A selection of loop patterns of size $6\times 6$. 
The length/pattern relation is: 
\textit{(length, pattern)} =  (20, a1..e5), (18, e6..f6), (16, f7..g4), (14, g5..g6), (16, g7..h3), (14, h4..h7), (12, h9 ..i5), (10, i6 ..i7), (8, i8...i9).
The following loops are equivalent under shift:  
(f7, f8, f9), (g1, g2), (g5, g6), (h4, h5), (h9, i1), (i2, i3, i4), (i6, i7), (i8, i9).
Uncovered cells are marked in red, if not on the boundary.
They have a distance (orthogonally or vertically) of 2 or 3 from a path cell. 
Uncovered cells with distance 3 can be observed at the borders of the patterns
a4, e9, g3, g5, h2, h7, i5, i7, i9. 
}
\label{ALL6x6}   
\end{figure}
%%%%%%%%%%%%%%%%%%%%%%%%%%%%%%%%%%%%%%%%% ALL6x6
%
%
The patterns are ordered with respect to 
\begin{enumerate}
	\item 
	the size of the rectangle  that encloses a loop, 
i.e. 
6$\times$6 (a1 .. e5), 
6$\times$5 (e6 ..  f6),
6$\times$4 (f7 .. g4), 
6$\times$3 (g5, g6), 
5$\times$5 (g7 .. h3), 
5$\times$5 (h4 .. h7), 
5$\times$3 (h8), 
4$\times$4 (h9 .. i5), 
4$\times$3 (i6, i7), 
3$\times$3 (i8, i9),
and 

\item
the number of inner corners.
\end{enumerate}

Some of the loops show the same structure but are shifted which results in a different 
covering configuration, often with uncovered cells (marked in red). 
The evolvable patterns depend on the rule variant being applied:

\begin{itemize}
	\item \textit{Rule0}. 
  All patterns can appear, but the problem is that the all-zero configuration 
  is also a fixed point as explained before. 
  
  \item \textit{Rule1}. 
  Noise is injected for uncovered cells which leads to either a configuration with no 
  uncovered cells or to a loop with toggling cells at distance 2 and/or 3.
  Some of the loops are destroyed (like g5) during the evolution because of noise propagation
  from toggling cells.
  
  \item \textit{Rule2}.
  Uncovered cells at a distance of 2 or 3 from a path cell are allowed.
  Now all patterns such as shown in Fig.~\ref{ALL6x6} can evolve and be stable.
  Now we can understand better the motivation for defining this rule variant:
  (i) a certain amount of uncovered cells shall be tolerated in order to 
  increase the set of finally stable solutions, and
  (ii) reduce the time to find a solution, because 
  the cardinality of the solution set is higher.

\end{itemize}

\newpage
%%%%%%%%%%%%%%%%%%%%%%%%%%%%%%%%%%%%%%%%%%%%%%%%%%%%%%
\subsection{Patterns of size $7\times7$}
%%%%%%%%%%%%%%%%%%%%%%%%%%%%%%%%%%%%%%%%%%%%%%%%%%%%%%

We can  only  show a selected subset  of all possible $7\times7$  loop patterns 
(Fig.~\ref{ALL7x7}).
%
%%%%%%%%%%%%%%%%%%%%%%%%%%%%%%%%%%%%%%%%% ALL7x7
\BEGINFIGURE
\includegraphics[width=0.9\textwidth]{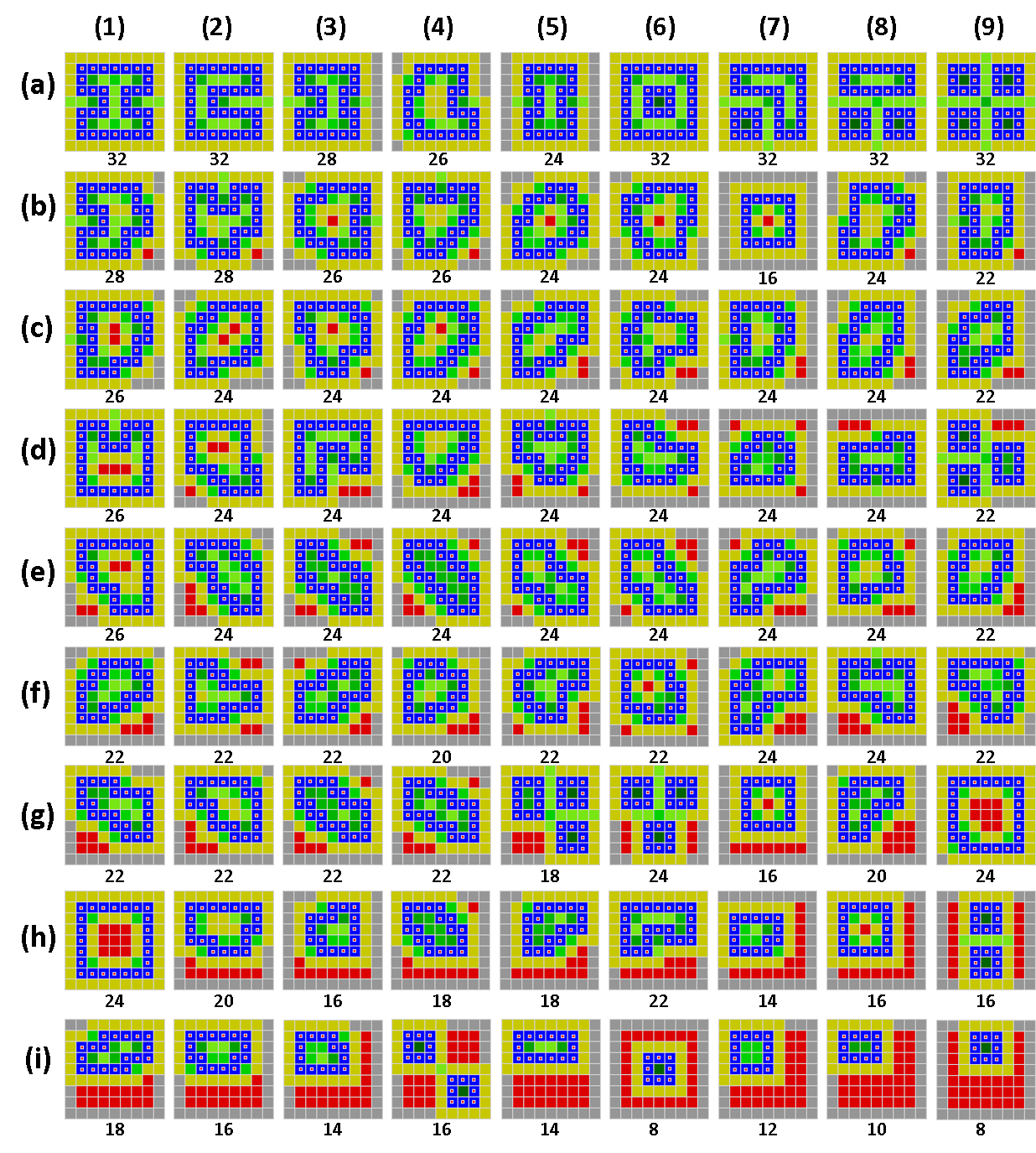}  %ok aber geringe qualität
\caption{%++ALL7x7++ 
A selection of loop patterns of size $7\times 7$. 
The loops are ordered by the (i) the number uncovered cells (in red),
and (ii) by the loop length (number of blue path cells).
The loop length is given under each pattern. 
}
\label{ALL7x7}   
\end{figure}
%%%%%%%%%%%%%%%%%%%%%%%%%%%%%%%%%%%%%%%%% ALL7x7
%
The patterns are ordered with respect to 
(i) the number uncovered cells and
(ii) the loop(s) length.
The class of patterns that can evolve  depends on the specific rule applied:

\begin{itemize}
	\item \textit{Rule0}. 
  All loop patterns can appear, inclusive the all-zero configuration.
  
  \item \textit{Rule1}. 
  Noise is injected for uncovered cells which leads to either a stable configuration with no 
  uncovered cells or to locally stable loops with unstable uncovered cells at a distance $\geq 2$.
  Some of the loops as shown in 
  Fig.~\ref{ALL7x7} 
  are only transients which are destroyed because of noise propagation from uncovered cells causing new template hits
  near a path or an already formed loop.

  \item \textit{Rule2}.
  Uncovered cells at a distance of 2 or 3 from a path cell are allowed and stable (e.g. see g5, h7, i7). 
  For uncovered cells at a larger distance noise is still injected which either drives 
  the evolution to a globally stable pattern,  
  or to a partially unstable pattern for cells at  larger distances. 
  Patterns with toggling cells at distance 4 are for example (i4, i5, i8, i9) which are relatively rare.
  Uncovered cells change their state because of noise injection
  which very often induces other template hits nearby. % and thereby state changing of the neighborhood. 
  
  \item \textit{Rule3}.   
  This rule variant is now applied for the first time.
  The motivation is that we now aim at patterns with no or a small number of uncovered cells and
  especially not as pairs at a distance of 2 and 3.     
  It destroys path cells if there are uncovered cells at a distance of 2 and 3. 
  The effect is that stable configurations with uncovered cells at a distance of 3 will no longer appear. 
  In this way patterns such as 
	(f7 -- f9, g1, g5, g8, h1, h5 -- h7, 
	i1 -- i5, i7 -- i9)
	are excluded from the set of stable patterns. 

  For comparison 300 patterns were evolved for \textit{Rule2} and \textit{3} as an experiment.
  All patterns were stable except two (with toggling cells) when \textit{Rule2} was used.
  The number of time steps on average $t_{avrg}$ to reach a stable (or toggling) pattern was computed,
  and also the average number of uncovered cells $u_{avrg}$.
  The result was  ($t_{avrg}= 33, u_{avrg}=3.3$) for \textit{Rule2}, and
  ($t_{avrg}= 103, u_{avrg}=2.5$) for \textit{Rule3}.
  If we were searching for totally covered patterns with no uncovered cell, then the success rate is higher for \textit{Rule3}
  (16 were found) than that for \textit{Rule2} (8 were found).   
  
  \item \textit{Rule4}.  
  This rule adds additional noise if the cover level is $v=8$.
  Such a cover level is only possible for the center of a $3\times 3$ square (Fig.~\ref{u4x4-1}).
  Therefore such local configurations
  as (a6 -- a9, d9, g5, g6, h9, i4, i6, i9)
  are destroyed and will not appear in the finally stable pattern.
  
\end{itemize}

%%%%%%%%%%%%%%%%%%%%%%%%%%%%%%%%%%%%%%%%%%%%%%%%%%%%%%
\subsection{Patterns of size $11\times11$}
%%%%%%%%%%%%%%%%%%%%%%%%%%%%%%%%%%%%%%%%%%%%%%%%%%%%%%

A selection of $11 \times 11$ pattern evolved with \textit{Rule4} is shown in 
Fig.~\ref{Selection11x11e23d23h8}, except the first one which is a closed
space-filling curve (SFC). %->
%
%%%%%%%%%%%%%%%%%%%%%%%%%%%%%%%%%%%%%%%%% Selection11x11e23d23h8
\BEGINFIGURE
%\sidecaption            %%max 7.8cm with sidecaptio

\includegraphics[width=0.9\textwidth]{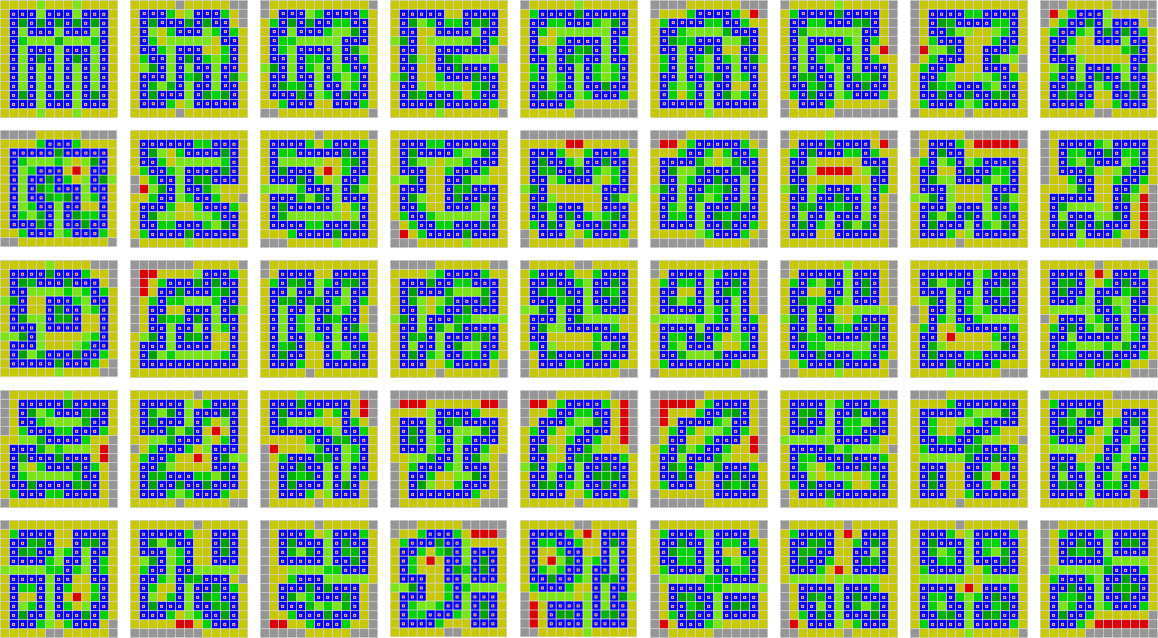}  %ok aber geringe qualität
\caption{%++Selection11x11e23d23h8++ 
A selection of loop patterns of size $11\times 11$ evolved by \textit{Rule4}. 
The loops are ordered by  the number of loops and the uncovered cells (in red).
}
\label{Selection11x11e23d23h8}   
\end{figure}
%%%%%%%%%%%%%%%%%%%%%%%%%%%%%%%%%%%%%%%%% Selection11x11e23d23h8
%
It was constructed and then showed stability. 
The probability to evolve such a highly symmetric and dense loop is relatively low. 
We can construct such SFC for fields of size $(4k-1)\times(4k-1)$ where $k=1, 2, ...$ .
It would be interesting to extend our approach aiming at SFC only.
The number of loops in the evolved pattern varies from one to four.
The maximum would be nine $3 \times 3$ loops, but such loops are not supported
by intention, because noise is injected if the cover level is 8 (a unique
attribute for the center of such a loop). 

The simulation snapshots of a sample
pattern evolution 
(Fig.~\ref{snapshot11x11})
show that transient loops are born which may die during the evolution
until a stable loop pattern is reached allover.

%%%%%%%%%%%%%%%%%%%%%%%%%%%%%%%%%%%%%%%%% snapshot11x11
\BEGINFIGURE
\includegraphics[width=0.9\textwidth]{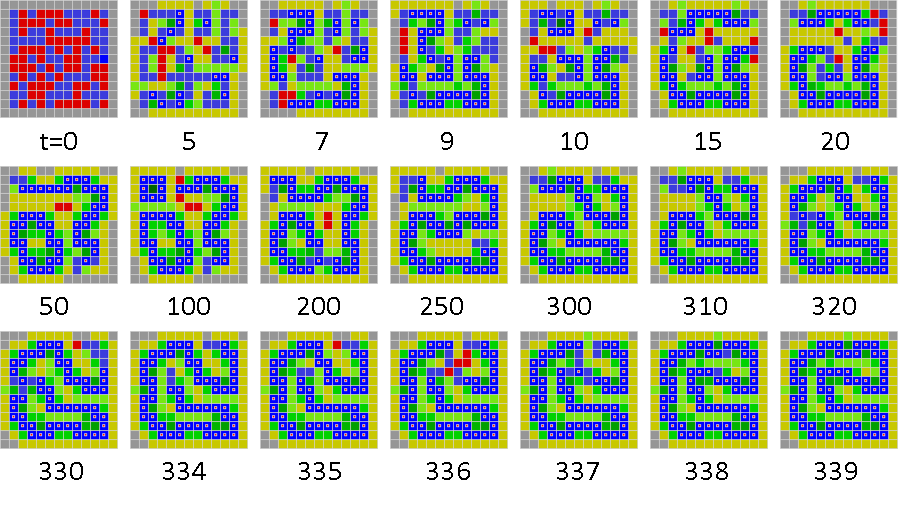}  %ok aber geringe qualität
\caption{%++snapshot11x11++ 
Snapshots of a $11\times 11$ 
 simulation using \textit{Rule4}, starting with a random configuration. 
}
\label{snapshot11x11}   
\end{figure}
%%%%%%%%%%%%%%%%%%%%%%%%%%%%%%%%%%%%%%%%% snapshot11x11

\newpage
%%%%%%%%%%%%%%%%%%%%%%%%%%%%%%%%%%%%%%%%%%%%%%%%%%%%%%
\subsection{Patterns of size $32\times32$}
%%%%%%%%%%%%%%%%%%%%%%%%%%%%%%%%%%%%%%%%%%%%%%%%%%%%%%

A pattern selection for a larger field of size $32\times32$
is shown is in  
Fig.~\ref{selection32x32} for \textit{Rule4}. %->
%
%
%%%%%%%%%%%%%%%%%%%%%%%%%%%%%%%%%%%%%%%%% Selection11x11e23d23h8
\BEGINFIGURE
\includegraphics[width=0.9\textwidth]{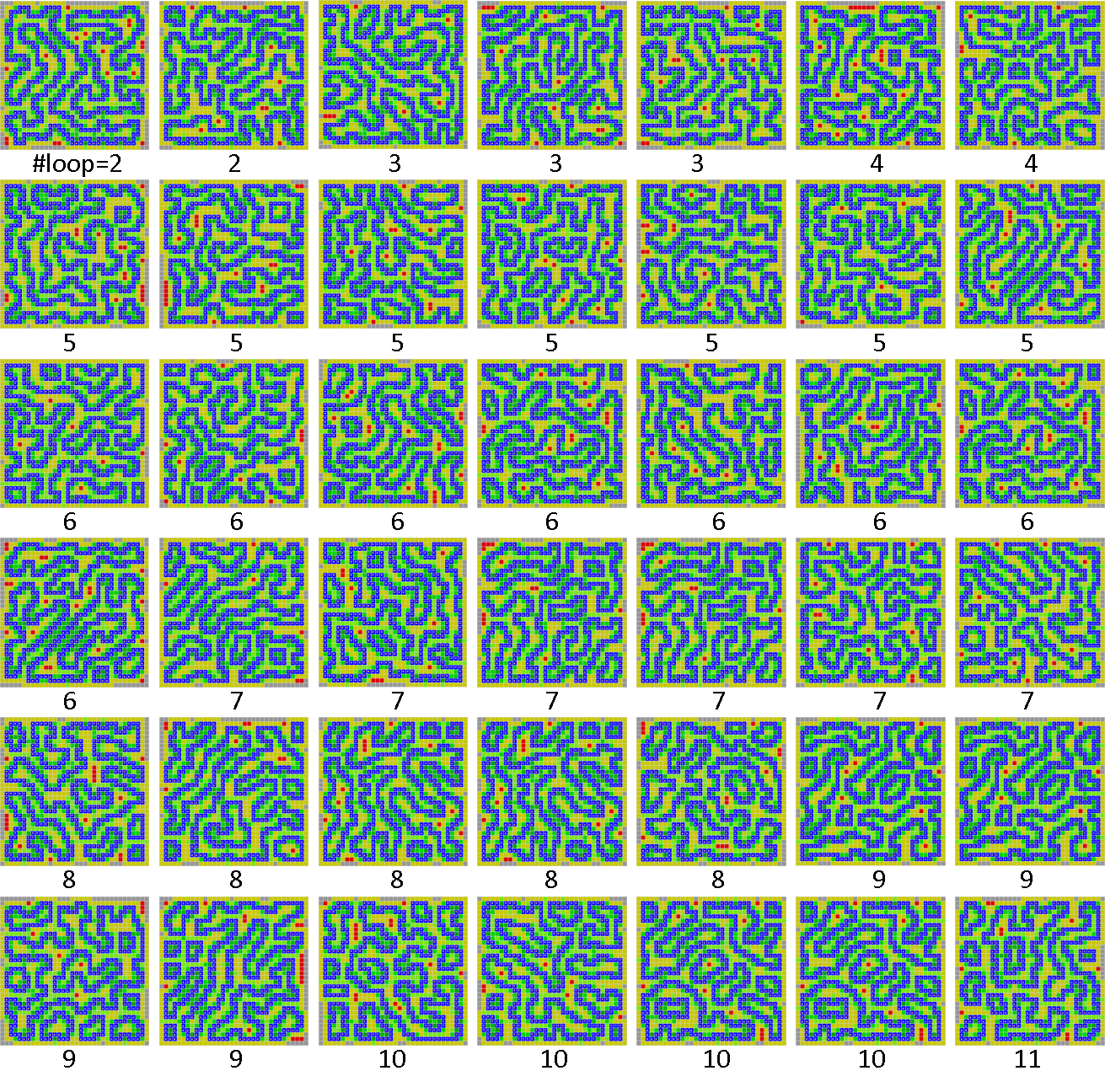}  %ok aber geringe qualität
\caption{%++selection32x32++ 
A selection of loop patterns of size $32\times 32$ evolved by \textit{Rule4}. 
The loops are ordered by the number of loops.
}
\label{selection32x32}   
\end{figure}
%%%%%%%%%%%%%%%%%%%%%%%%%%%%%%%%%%%%%%%%% selection32x32
%
%
The number of loops in the patterns vary from 2 to 11.
No loop pattern was found with one loop only, because the
probability is very low. 
It would be possible to find this probability by extensive simulations,
and it would be interesting to find it by formal methods.
Also loop patterns with a maximum number of loops do not appear easily. 
We could for instance construct a pattern with 64 $3 \times 3$ square
loops, although they are excluded for \textit{Rule4}.
Thus the question arises:
What is the maximum number of loops if $3 \times 3$ loops are forbidden. 
On idea is to place $3 \times 4$ loops only.
This could be formulated as a separate task, using  $5 \times 6$ tiles
with a $3 \times 4$ kernel of one-pixels. 
One could then see this problem as a generalized rectangle tiling problem
where rectangles of any (small) size are used.
For that problem we see a relation to the domino tiling problem
\cite{2023-Hoffmann-mdpi-loop-pattern}
\cite{Hoffmann:Deserable-Seredynski-2021a-JSup-A cellular automata rule placing a maximal number of dominoes in the square and diamond}
\cite{2023-Dominique-dominoes-arxiv}
already studied.   

%%%%%%%%%%%%%%%%%%%%%%%%%%%%%%%%%%%%%%%%%%%%%%%%%%%%%%
\subsection{Some dependencies}
%%%%%%%%%%%%%%%%%%%%%%%%%%%%%%%%%%%%%%%%%%%%%%%%%%%%%%

It would be of interest to study certain dependencies 
and distributions theoretically and/or  practically through 
extensive simulations.
Parameters or variables are for instance:
the field size, 
the convergence time,
the number of uncovered cells,
or the number, shape, length, area or density of  loops.
As we are here not able to investigate such dependencies in detail, we
just want to give a few results 
that depend on the field size obtained by simulation. 

Simulation experiments with \textit{Rule4} for different $n$ were conducted in order to find out how the
number of \textit{time-steps},
the number of \textit{uncovered cells},
and the \textit{loop density} depend on the field size. 
For $n=4$ to $n=15$ the number of runs was 300, and only 100 for $n=16$ to $n=20$
because of limited computing time.
We can observe a super-linear increase for the 
number of micro time-steps time per cell 
(Fig.~\ref{TimeComplexity}),  
and we guess that it is exponential although it needs to be proved. %->
%
%%%%%%%%%%%%%%%%%%%%%%%%%%%%%%%%%%%%%%%%% TimeComplexity --TimeVSn
\BEGINFIGURE
\includegraphics[width=0.5\textwidth]{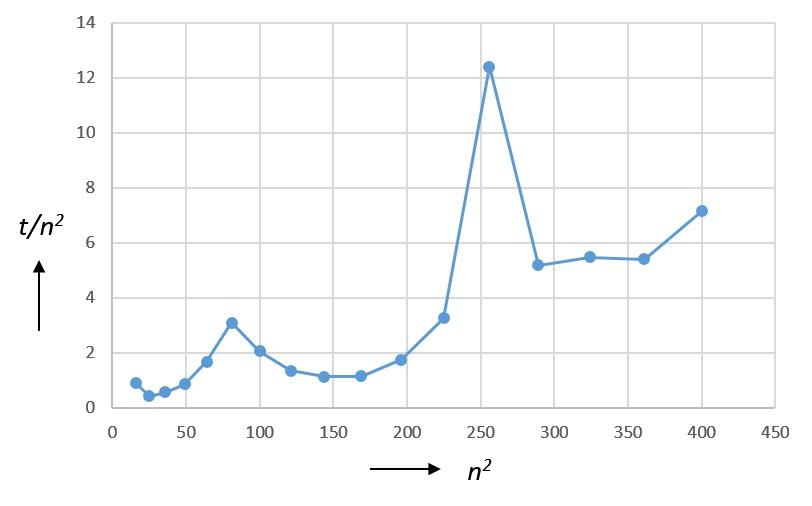}  
\caption{%++TimeComplexity++ 
The average number of micro time-steps per cell to evolve a stable pattern depending on the 
number $n^2$ of cells.
Rule 4 was applied. 
%The dotted trend line is an exponential approximation.
}
\label{TimeComplexity}   
\end{figure}
%%%%%%%%%%%%%%%%%%%%%%%%%%%%%%%%%%%%%%%%% TimeComplexity --TimeVSn
%
The number of uncovered cells and the loop density (number of one-cells divided by $n^2$)
is depicted in
Fig.~\ref{LoopDensity}. 
%
%
%%%%%%%%%%%%%%%%%%%%%%%%%%%%%%%%%%%%%%%%% LoopDensity
\BEGINFIGURE
\includegraphics[width=0.45\textwidth]{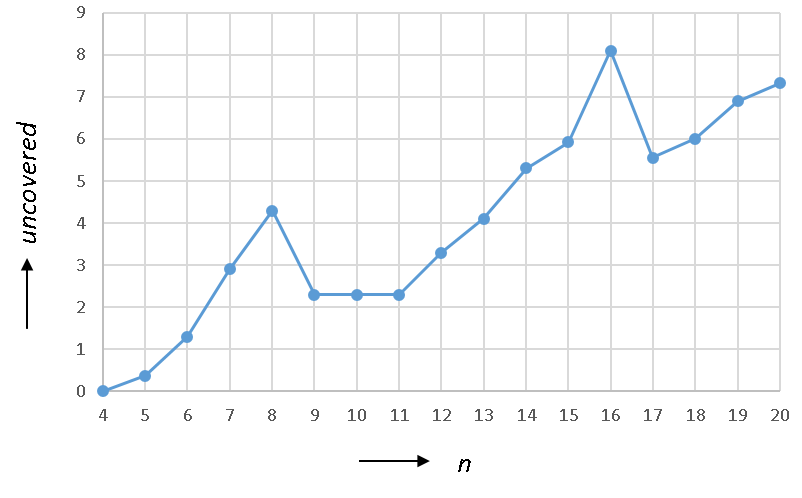}  
~~\includegraphics[width=0.45\textwidth]{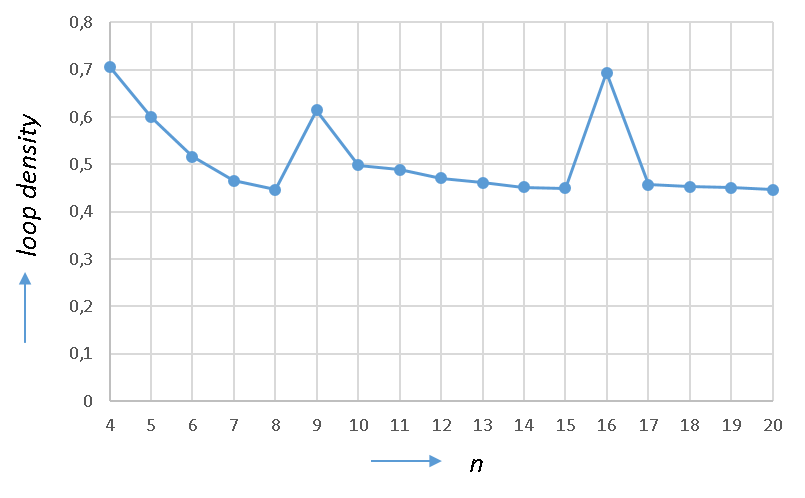}  
\caption{%+UncoveredVSn+LoopDensity++ 
(Left) The number of uncovered cells.  % vs $n$.
(Right) The loop density (number of loop path cells divided by $n^2)$.
}
\label{LoopDensity}   
\end{figure}
%%%%%%%%%%%%%%%%%%%%%%%%%%%%%%%%%%%%%%%%% LoopDensity
%
%
%\begin{bf}  %%%%%%%%%%%%%%%%%%%%%%%%%%%%%%%%%%%%%%%%%%%%%% begin NEW
%\color{red} %%%%%%%%%%%%%%%%%%%%%%%%%%%%%
There are exceptional peaks for $n=9$  and $n=16$
in both the number of micro time-steps and the loop density.
%(resp. $n=8$ for the uncovered cells)
A general explanation is that we have it to do  with a complex combinatorial problem of placing
certain loop structures in a given field,
tolerating a certain range of distance between them,
and taking into account covering constraints.
%Therefore the possible solutions may be more difficult to reach for $n=16$ than for $n=15, 17$. 
%Therefore the number and complexity of possible solutions does not increase with $n$ in a simple way,
%i.e. there may be more complex and therefore difficult to reach solutions for $n=16$ than for $n=15, 17$.
This issue is not really clear and needs further investigations in order to find out if
these singularities are inherent to the problem or if they are artifacts caused by the CA algorithm. 

Other simulations with 100 runs (experiments)  were performed for $n=15,16,17$ using \textit{Rule4} 
in order to get an impression how the number of loops are distributed.
The result was
\textit{(occurrence, number of loops)} = 

\MYTAB{5mm}{
(26, 1), \underline{(46, 2)}, (21, 3), (7, 4)  ~~~for $n=15$; 

(6, 1),  (24, 2), \underline{(30, 3)}, (26, 4), (10, 5), (3, 6), (1, 7)  ~~~for $n=16$;

(15, 1),  \underline{(28, 2)}, (24, 3), (17, 4), (12, 5), (4, 6)  ~~~for $n=17$.
}

The average number of loops was $2.09, ~3.23, ~2.95$ ~~~for $n=15,16,17$.

Although we need more experiments for a statistical evidence we can observe that the occurrence of
two (or three) loops is most likely, and patterns with increasing more loops 
become more and more unlikely.

\COMMENT{
Another experiment was performed with \textit{Rule3} on a $11 \times 11$ field with 300 runs
in order to get an impression for the distribution of the number of loops.
The result was
\textit{(number of loops, occurrence)} = (1, 65), (2, 161), (3, 60), (4, 14). 
No loops were evolved with a loop length of 5 to 9 (maximal nine $3 \times 3$ loops can be placed).
For this example it means that the probability to evolve a pattern with a high number
of (small) loops is very low.  
}

%%%%%%%%%%%%%%%%%%%%%%%%%%%%%%%%%%%%%%%%%%%%%%%%%%%%%%
\section{Summary and open issues}
\label{S5}
%%%%%%%%%%%%%%%%%%%%%%%%%%%%%%%%%%%%%%%%%%%%%%%%%%%%%%
%%%%%%%%%%%%%%%%%%%%%%%%%%%%%%%%%%%%%%%%%%%%%%%%%%%%%%
\subsection{Summary}
%%%%%%%%%%%%%%%%%%%%%%%%%%%%%%%%%%%%%%%%%%%%%%%%%%%%%% 

We have designed a CA rule that is able to evolve stable loop patterns. 
The rule is probabilistic and it is simulated under asynchronous updating. 
It tests the CA against a set of templates (local patterns) for matching.
The templates are derived from a set of tiles which can overlap and finally form the pattern.
The defined tiles are corners and line segments.
Noise is injected as long as certain conditions are not fulfilled
and the evolving loop pattern is not stable. 
\NEW{
The most important condition, the \textit{loop path condition}, is realized by an overlap level of three for every  one-cell.  
}

A template hit is registered if a template matches. 
Two kind of hits are distinguished, 
$h_0$ if the state (or the adjusted state) is zero,
and $h_1$ if the state (or the adjusted state) is one. 
If there are several concurrent $h_0$ and $h_1$ hits,
then the state adjustment is random. 

For statistics and visualization it is useful to compute the cover level.
It gives the number of overlaid pixels from different tiles. 
It can easily computed by an extra sweep over the whole field checking for tile matches.
During the sweep the position of the centers of the matching tiles can be stored and marked. 
 
A difficulty during the design and simulations was to make the rule simple and reliable.
Five rule variants % specialized rules
of the general CA rule were defined depending on certain activated conditions. 
The most simple \textit{Rule0} can easily produce any loop patterns but can also get stuck in the 
all-zero configuration. 
\textit{Rule1} avoids the all-zero configuration but the pattern may only be stable in parts. 
\textit{Rule2} turns unstable uncovered cells at distance 2 or 3 into stable uncovered ones. 
\textit{Rule3} evolves stable pattern and allows uncovered cells at distance 2.
\textit{Rule4} forbids cover level 8 which is intended to destroy small loops of size $3\times 3$.

For this mainly experimental work, 
the minimal loop pattern size is $3 \times 3$, and there is no limit
for large sizes.
For large sizes, we expect a mixture of different loop shapes and loop sizes
as in Figure~\ref{selection32x32}.
The CA rules \textit{Rule3} and \textit{Rule4} always generated stable loop patterns so far. 
The computation time increases exponentially in the number of cells, 
shown for experiments with \textit{Rule3}.
%In the simulation, rectangular fields can be handled too. 

%\newpage
%%%%%%%%%%%%%%%%%%%%%%%%%%%%%%%%%%%%%%%%%%%%%%%
\subsection{Open issues}
%%%%%%%%%%%%%%%%%%%%%%%%%%%%%%%%%%%%%%%%%%%%%%
This approach of forming patterns with overlapping tile raises a lot of open issues
for further research. 

\begin{itemize}
\item
The used settings of the probabilities for the decisions in the CA rule were good working,
but there is potential for optimizing them in order to minimize the
average time for finding a stable pattern. 
The computation time can also be reduced by parallel processing (parallel threads, parallel hardware).
Nevertheless it %the computation time 
is increasing exponentially with the number of cells
as already tested for domino patterns in 
\cite{Hoffmann:Deserable-Seredynski-2021a-JSup-A cellular automata rule placing a maximal number of dominoes in the square and diamond}.
The question is if there are other algorithmic methods (like divide and conquer)
which could reduce the time complexity.
  
\item  
The convergence and limits of this approach 
should be proved and further evaluated. 
Can it be proven that the presented CA rule always 
(especially for large field sizes) produces stable loop patterns?

\item
Can this approach easily be extended to higher dimensions or more states (colors)?% or other boundary conditions?

\item
Can synchronous updating be used, and how has the rule to be modified?

\item
How can the rule or simulation algorithm be changed in order to prioritize certain tiles.
One way is to store the hits separately for each tile and then 
decide upon the level of noise injection. 

\item 
What is the relation between the loop types (e.g. the number and sequence of corners)
and size of the cell field?
%How are the loop types with certain properties distributed?
What are the probabilities that certain loop types
are evolved?    
% What are (all) possible loop patterns depending on the  size of a field, and  what is the distribution of different loop types with certain properties?
% A special question could be: How many loops (optionally restricted to a certain type)  can maximal be placed in  a certain~field.

\item
Is it possible to evolve closed space filling curves only or with a high probability?

\item
How depends the patterns structure on other or cyclic boundary conditions, 
or sub-areas that are defined by certain constant or variable sub-patterns. 

\item
Can this approach be generalized in order to be applied to other optimization or computational tasks?
Maybe this approach can be modified into a metaheuristic for optimization problems in general. 

\item 
The method is related to the vertex cover problem. This relation could be studied in more detail. 

\item 
How can optimal loops be generated on the basis of local conditions only
when optimality depends on a global measure.
Possible parameters for a global measure are:
The loop density / the space between loops
(In our former work
\cite{Hoffmann:Deserable-Seredynski-2021a-JSup-A cellular automata rule placing a maximal number of dominoes in the square and diamond}
\cite{Hoffmann:Deserable-Seredynski-pact-2021b-Minimal Covering of the Space by Domino Tiles}
the space between dominoes could be minimized / maximized by injecting noise
depending on the number of template hits.),    
or the number, length, or type of the loops.    

\item
How close is this approach to pattern formation in nature?
What are important areas of applications?
In this article, the evolution of patterns occurred randomly across the entire field, without nor interactions nor preference mechanism. In the case of physical or biological systems, loops are usually formed under action of physical or biological conditions, such as the stress field in the case of cracks, for example. In this context, the results presented here can be a starting point for studying what types of interactions/preferences affect the process of loop formation and what are their qualitative and quantitative effects. Consequently, this can help capture the universal aspects of loop formation, which resonates with the research approach for the reticulated pattern \cite{A work on reticulated patterns}.

\item
How sensitive are the evolved patterns against the initial conditions?
First experiments have shown that the initial pattern does not have
a strong influence except when the initial pattern is close to a stable loop pattern. 
The initial hit values have also to be taken into account. 

\item
How sensitive is the evolution against the order in which the cells are selected for updating?
First experiments have shown that this sensitivity is low. 
Special  deterministic updating orders could be considered. 

\item
The probabilities that certain loops evolve in a certain time could be 
treated theoretically. The distribution of certain loops could be
attacked from the combinatorial point of view.  

\end{itemize}

In conclusion, we found a method to generate plane 
loop patterns under local conditions only
but many issues remain open for further research.

%%%%%%%%%%%%%%%%%%%%%%%%%%%%%%%%%%%%%%%%%%%%%%%%%%%%%%%%%%%%%%%%%%%%%%%%%%references
%%%%%%%%%%%%%%%%%%%%%%%%%%%%%%%%%%%%%%%%%%%%%%%%%%%%%%%%%%%%%%%%%%%%%%%%%%references
%\newpage

%%%%%%%%%%%%%%%%%%%%%%%%%%%%%%%%%%%%%%%%%%%%%%%%%%%%%%%%%%%%%%%%%%%%%%%%%%references
\end{document}